\documentclass[fleqn,usenatbib]{mnras}

\usepackage{newtxtext,newtxmath}

\usepackage[T1]{fontenc}

\DeclareRobustCommand{\VAN}[3]{#2}
\let\VANthebibliography\thebibliography
\def\thebibliography{\DeclareRobustCommand{\VAN}[3]{##3}\VANthebibliography}

\usepackage{graphicx}	
\usepackage{amsmath}	
\usepackage{mathtools}
\usepackage{bigints}
\usepackage{soul}

\title[Effect of streaming instability on disc emission]{{The effect of the streaming instability on protoplanetary disc emission at millimetre wavelengths}}

\author[Scardoni, Booth, Clarke]{Chiara E. Scardoni,$^{1}$\thanks{E-mail: ces204@ast.cam.ac.uk}, Richard A. Booth$^{1,2}$, Cathie J. Clarke$^{1}$\\
$^{1}$Institute of Astronomy, University of Cambridge, Madingley Road, Cambridge, CB3 0HA, United Kingdom\\
$^{2}$Astrophysics Group, Imperial College London, Blackett Laboratory, Prince Consort Road, London SW7 2AZ, United Kingdom
}

\date{Accepted XXX. Received YYY; in original form ZZZ}

\pubyear{}

\begin{document}
\label{firstpage}
\pagerange{\pageref{firstpage}--\pageref{lastpage}}
\maketitle

\begin{abstract}
In this paper, we investigate whether overdensity formation via streaming instability is consistent with recent multi-wavelength ALMA observations in the Lupus star forming region. We simulate the local action of streaming instability in 2D using the code {\sc Athena}, and examine the radiative properties at mm wavelengths of the resulting clumpy dust distribution by focusing on two observable quantities: the optically thick fraction $ff$ (in ALMA band 6) and the spectral index $\alpha$ (in bands 3-7). By comparing the simulated  distribution in the $ff-\alpha$ plane before and after the action of streaming instability, we observe that clump formation causes $ff$ to drop, because of the suppression of emission from grains that end up in optically thick clumps.
$\alpha$, instead, can either increase or decline after the action of streaming instability; we use a simple toy model to demonstrate that this behaviour depends on the sizes of the grains whose emission is suppressed by being incorporated in optically thick clumps. In particular, the sign of evolution of $\alpha$ depends on whether grains near the opacity maximum at a few tenths of a mm end up in clumps.
By comparing the simulation distributions before/after clump formation to the data distribution, we note that the action of streaming instability drives simulations towards the area of the plane where the data are located. We furthermore demonstrate that this behaviour is replicated in integrated disc models provided that the instability is operative over a region of the disc that contributes significantly to the total mm flux.
\end{abstract}




\section{Introduction}
According to core accretion theory, planets form through dust growth from initial $\mu\rm m$-sized grains up to the size of a planet \citep{Safronov&Zvjagina1969}. Since a substantial change in the grain mass and size is required in this process, different physical processes are expected to happen during dust growth; therefore, the process is often divided into three main stages (e.g., \citealt{Lissauer1993,Papaloizou&Terquem2006,Armitage2007_BOOK}): grain growth, where the dust grains grow from $\mu$m to cm, mainly through collision and sticking processes \citep{Dominik&Tielens1997,Birnstiel+2012,Garaud+2013,Dominik+2016}; planetesimal formation, from cm-sized dust grains to km-sized planetesimals; protoplanet formation, which leads to spherical objects of $\sim10^3$ km, that could be either rocky planets or gaseous planet cores (see, for example, \citealt{Safronov&Zvjagina1969,Kokubo&Ida1996,Rafikov2003,Ormel+2010,Kobayashi+2016,Kobayashi&Tanaka2018}).

The planetesimal formation stage represents a critical step in the growth process because the formation of km-sized objects is hampered by the so-called ``metre-sized barrier'' or ``radial drift barrier'' \citep{Adachi+1976,Weidenschilling1977,Takeuchi&Lin2002,Brauer+2008,Pinte&Laibe2014}. The aerodynamic interaction between dust and gas in protoplanetary discs, in fact, causes dust particles to lose angular momentum and to drift inwards. How rapidly the particles drift depends on the coupling between the gas and the dust, which is usually measured through the so-called Stokes number, $\tau_{\rm s}$. Since $\tau_{\rm s}$ is proportional to the grain size, $a$, we can relate the radial drift to $a$, and it is possible to show that for grains characterised by $a\sim {\rm cm-m}$ drift can be as fast as $t_{\rm drift}\sim 100\ \rm yr$  at $1\ \rm AU$ from the star \citep[for example]{Armitage2007_BOOK,Birnstiel+2016}. Consequently, unless other mechanisms interfere with the inward drift, the $\sim$cm-m sized grains are soon lost and they are no longer available to form planetesimals.

A potential way to avoid the metre-sized barrier is dust concentration in overdensities that rapidly collapse due to self gravity \citep{Johansen+2007}, and the streaming instability \citep{Youdin&Goodman2005} is a popular idea for how this might happen \citep{Johansen+2007,Youdin&Johansen2007,Johansen&Youdin2007}. Like  radial drift,  the streaming instability is caused by the interaction between dust and gas components within the disc, and, more precisely, by the backreaction of dust on the gas. When exerted by partially coupled particles ($\tau_{\rm s}\sim 1$), the dust backreaction is strong enough that it can locally trigger on the disc midplane a powerful hydrodynamic instability (the streaming instability, \citealt{Youdin&Goodman2005}), which, in appropriate conditions, promotes fast particle clumping, forming dust overdensity regions (\citealt{Youdin&Johansen2007,Carrera+2015,Yang+2017}). Thus, streaming instability might be the key to solving the planetesimal formation issue: on one hand, the formed clumps would interact in a different way with the gas and this would slow down the drift velocity; on the other hand, local overdensities would facilitate the gravitational collapse and then planetesimal formation \citep{Johansen+2007}.

Clump formation through streaming instability can qualitatively be described as follows. The backreaction on the gas causes the gas component to orbit faster, and therefore it reduces the difference in azimuthal velocity between the gas and the dust, thus the headwind on the particles becomes weaker \citep{Youdin&Goodman2005,Johansen&Youdin2007,Squire&Hopkins2020}. If a dust overdensity were present, it would perturb the system equilibrium, causing a stronger backreaction, and therefore a reduced radial drift (which is determined by the difference in velocity between the two components). The system therefore faces  an instability, as the initial dust overdensity increases more and more due to the new material drifting inward from outer orbits and stopping in the clump (\citealt{Johansen&Youdin2007} described this effect as a ``traffic jam''); the consequence is an exponential growth of clumps, whose density can become as high as $10^3$ times the gas density, $\rho_{\rm g}$ (e.g, \citealt{Youdin&Johansen2007,Bai&Stone2010-2,Bai&Stone2010}). At the same time some particles are lost from the inner radius of the clumps, due to radial drift; therefore at some point a steady state is reached and the exponential growth saturates. The development of such exponentially growing instabilities is expected to be easier for marginally coupled particles \citep{Johansen&Youdin2007}, because their backreaction is stronger (the fastest radial drift is expected to occur for $\tau_{\rm s}\sim 1$).

The ability of streaming instability to form dust clumps depends on the local characteristics of the system, and in particular on the following three parameters: the grain Stokes number $\tau_{\rm s}$, which regulates the gas-dust interaction; the pressure support, $\Pi$, which gives rise to  the relative velocity and thus ultimately generates the streaming instability; and the local dust to gas mass ratio, $Z$, which must exceed a certain threshold in order to create dust clumps. From the theoretical point of view, several studies focused on exploring this parameter space, in order to find which combinations of $(\tau_{\rm s}, \Pi, Z)$ can trigger dust clumping via streaming instability. \citet{Bai&Stone2010-2,Bai&Stone2010} showed that, for a given grain size population, the stronger the pressure support, the higher is the critical $Z$ to have streaming instability; whereas, for a given $Z$, smaller values of $\Pi$ allow smaller particles to form clumps. More recently, \citet{Carrera+2015} and \citet{Yang+2017} studied in detail the critical $Z$ required to trigger streaming instability {for fixed $\tau_{\rm s}$ (single grain population) and $\Pi$}, identifying the regions of the $Z-\tau_{\rm s}$ plane where streaming instability is expected to occur. Several studies also focused on the properties of streaming instability in multi-species simulations, both in the linear instability phase \citep{Laibe&Price2014,Drazkowska+2016,Krapp+2019,Zhu&Yang2020} and in the non-linear phase \citep{Bai&Stone2010,Schaffer+2018}. They found that, under certain conditions, the efficiency of linear growth can be increased/decreased when multiple grain species are considered with respect to the single grain models.
As well as the physical properties of streaming instability, recent works studied in detail the influence of numerical parameters (size of the simulation box, resolution, etc.), algorithms and boundary conditions on streaming instability simulations \citep{Yang&Johansen2014,Li+2018}.

From an observational perspective, the recent Disk Substructures at High Angular Resolution Project (DSHARP) survey \citep{Andrews2018}, conducted with the Atacama Large Millimeter Array (ALMA), provided high resolution data for $20$ protoplanetary discs, finding in most of them interesting substructures, such as rings, spirals, gaps, etc. These observations are allowing a more detailed analysis of the processes operating in protoplanetary discs, including planetesimal formation, thereby deepening our knowledge of these systems. Among  other results, it has been found that the rings are marginally optically thick \citep{Huang+2018,Dullemond+2018}; this feature may be a mere coincidence, but it could also be the consequence of a common mechanism happening in rings, e.g. clump formation via streaming instability \citep{Stammler2019}.

Furthermore, \citet{Tazzari+2020a,Tazzari+2020b} recently presented a new ALMA survey at 3 mm in Lupus star forming regions. By combining these results to the previous surveys at 0.9 mm and 1.3 mm \citep{Ansdell+2016,Ansdell+2018}, they were able to study in detail the optical properties of observed systems and to extract important information about disc radii, the size-luminosity relation, the system optical depth, etc. Focusing on the spectral index and the optically thick fraction\footnote{Defined as the ratio of the flux to that of an optically thick disc of the same size: see section \ref{subsec:TheoComparisonObservations}}, they noticed that the data  concentrate in the region of spectral index $\sim 2.4-3$ and with a range of values of the optically thick fraction, presenting  a number of potential models that might explain the observed distribution in this plane.

In this context, the aim of this paper is to investigate whether the formation of optically thick substructures through the action of streaming instability is consistent with the observation by \citet{Tazzari+2020a,Tazzari+2020b}. We therefore simulate systems where streaming instability is present (taking advantage of the broad parameter analysis performed in the studies mentioned above) and analyse their optical properties after the formation of dust clumps. We then compute the two observable quantities considered by \citet{Tazzari+2020a,Tazzari+2020b} -- the optically thick fraction and the spectral index -- in order to perform a comparison between the data and the final distribution from our simulations.

The paper is organised as follows: in \sectionautorefname~\ref{sec:dustgasInteraction} we briefly summarise the main parameters and equations describing streaming instability; in \sectionautorefname~\ref{sec:methods} we describe the simulations performed and the method applied to analyse their optical properties; \sectionautorefname~\ref{sec:density} illustrates the process of clump formation in our simulations; the explanation of the consequences of the presence of substructures on optical properties is covered by \sectionautorefname~\ref{sec:LocalModelSI}, where we also introduce a toy model which mimics the action of streaming instability, and we compare the local results to the {{ disc-averaged observations of }}\citet{Tazzari+2020b}; in \sectionautorefname~\ref{sec:IntDiscModel} {{we define}} an integrated disc model and its optical properties are {{compared with the foregoing local results}}; finally, in \sectionautorefname~\ref{sec:discussion} we discuss the influence of the main model parameters and grain composition on the final results.

\section{Dust-gas interaction and streaming instability}
\label{sec:dustgasInteraction}
Protoplanetary discs contain both dusty and gaseous material, in an overall initial dust to gas ratio of about $1:100$. Both the components orbit around the central star, but their azimuthal velocities $v_{\phi}$ are moderately different between each other; in fact, $v_{\phi}$ is the result of the balance of gravitational, pressure and centrifugal forces in the case of gas; whereas the dust is subject only to gravity and centrifugal force, and the absence of pressure force makes the dust azimuthal velocity slightly faster than that of the gas. This results in an aerodynamic interaction between the two components, exerted through drag forces, which (together with the consequent backreaction) plays the main role in triggering the streaming instability.

In this paper, we consider the Epstein drag force (valid for particles smaller than their main free path, \citealt{Epstein1924})
\begin{equation}
    F_{\rm D} = \frac{4\pi}{3}\rho_{\rm g} a^2vv_{\rm th},
\end{equation}
where $a$ is the particle radius, $\rho_{\rm g}$ is the gas density and $v_{\rm th}$ is the gas thermal velocity. The ratio between the particle momentum and the drag force provides the interaction timescale between the gas and the dust grains of different sizes
\begin{equation}
	t_{\rm s}= \sqrt{\frac{\pi}{8}}\frac{\rho_{\rm p} a}{\rho_{g} v_{\rm th}},
\end{equation}
which can be conveniently expressed in units of the dynamical timescale $\Omega^{-1}$
\begin{equation}
	\tau_{\rm s}=t_{\rm s}\Omega,
	\label{eq:StGen}
\end{equation}
obtaining the dimensionless stopping time, usually called Stokes number, which provides information about the coupling between gas and dust: $\tau_{\rm s}\ll 1$ means that the timescale for interaction is significantly smaller than the dynamical time, thus the particles, at first approximation, follow the gas motion; on the contrary, if $\tau_{\rm s}\gg 1$, the dust-gas coupling is weak and the particle motion is barely affected by the gas.

\equationautorefname~\eqref{eq:StGen} can also be reversed to compute the grain size corresponding to a given Stokes number
\begin{equation}
    a=\frac{2}{\pi}\frac{\tau_{\rm s}\Sigma_{\rm g}}{\rho_{\rm p}},
    \label{eq:apart}
\end{equation}
where the $v_{\rm th}$ is written in terms of the isothermal sound speed $c_{\rm s}\sim \Omega_{\rm k}H_{\rm g}$, and $\rho_{\rm g}=\rho_0\exp[-z^2/(2H_{\rm g}^2)]=\Sigma_{\rm g}/(\sqrt{2\pi}H_{\rm g})$ is the gaussian profile assumed for the gas density ($\Sigma_{\rm g}$ is the vertically integrated gas density).

The pressure gradient, which is the main cause of drag between dust and gas, is often characterised by the so-called pressure support parameter, defined as
\begin{equation}
	\Pi = \eta \frac{v_{\rm k}}{c_{\rm s}} = \eta \left(\frac{H_{\rm g}}{R}\right)^{-1},
	\label{eq:press_gradient}
\end{equation}
where $H_{\rm g}$ is the gas scale height, $v_{\rm k}$ is the keplerian velocity, $c_{\rm s}$ is the sound speed, and $\eta$ is related to the pressure gradient
\begin{equation}
    \eta = \frac{1}{2}\left(\frac{H_{\rm g}}{R}\right)^2\frac{\partial \log P}{\partial \log R}.
\end{equation}
Finally, we introduce the dust to gas ratio (also called metallicity), defined as the ratio between the dust and the gas densities ($\Sigma_{\rm d}$ and $\Sigma_{\rm g}$, respectively)
\begin{equation}
    Z=\frac{\Sigma_{\rm d}}{\Sigma_{\rm g}},
\end{equation}
which is required to be high in order to trigger streaming instability. Note that the requirement of high $Z$ is just a local requirement, thus it is not in tension with the fact that the global dust to gas ratio is of the order of $1:100$.

\citet{Bai&Stone2010_athenaparticles} implemented in \textsc{Athena} the dust-gas equations, written in a local reference frame located at a fixed radius $R_0$ and rotating at angular velocity $\Omega$ (non-inertial reference frame). The continuity equation for the gas is
\begin{equation}
    \frac{\partial\rho_{\rm g}}{\partial t}+\nabla{\rho_{\rm g}\mathbf{u_{\rm g}}}=0,
\end{equation}
where $\rho_{\rm g}$ and $\mathbf{u_{\rm g}}$ are the gas density and velocity, respectively; the Euler equation for the gas is
\begin{align}
    \frac{\partial\rho_{\rm g}\mathbf{u_{\rm g}}}{\partial t}&+\nabla{\left(\rho_{\rm g}\mathbf{u_{\rm g}}\mathbf{u_{\rm g}}+P\mathbf{I}\right)}=\nonumber\\
    &=\rho_{\rm g}\left(2\mathbf{u_{\rm g}}\times\boldsymbol{\Omega }+3\boldsymbol{\Omega }^2 x \mathbf{\hat{x}}-\boldsymbol{\Omega }^2 z \mathbf{\hat{z}}+\sum_k \epsilon_k \frac{\mathbf{v_k} -\mathbf{u_{\rm g}}}{t_{{\rm s},k}}\right),
    \label{eq:EulerGas}
\end{align}
where $P$ is the pressure, $\epsilon_{k}$ is the dust to gas ratio (for particle species $k$), $t_{s,k}$ is the stopping time of $k$ particle species, $\sigma_k$ and $\mathbf{v_k}$ are the particle local mass density and velocity, respectively. The left-hand side includes the momentum time derivative and the effects due to advection and pressure gradient; the right-hand side includes the action of Coriolis force (first term), radial tidal momentum (second term), vertical gravity (third term) and backreaction (fourth term). The i-particle equation of motion is
\begin{equation}
    \frac{{\rm d}\mathbf{v_{i}}}{{\rm d} t}=-2\eta v_{\rm k}\boldsymbol{\Omega}\mathbf{\hat{x}}+ 2\mathbf{v_{i}}\times\boldsymbol{\Omega}+3\boldsymbol{\Omega }^2 x_i \hat{x}-\boldsymbol{\Omega}^2 z_i \hat{z}- \frac{\mathbf{v_i} -\mathbf{u_{\rm g}}}{t_{{\rm s},k}},
    \label{eq:particles}
\end{equation}
where the first term on the right-hand side is the effect of the pressure gradient, and all the other terms corresponds to those in \equationautorefname~\ref{eq:EulerGas}.

We have neglected the self-gravity of the dust in these simulations. Since, as we show later, the clumps formed in our simulations without self-gravity already contribute negligibly to the total flux of the disc, the inclusion of self-gravity, which acts to further enhance clumping, is not expected to have a significant effect on the observed properties.

We also neglected the effects of turbulence, as is common in streaming instability simulations \citep{Liu&Ji2020}. Turbulence is generally expected to reduce particle clumping via streaming instability \citep{Umurhan+2020,Gole+2020}, therefore the usual no-turbulence assumption corresponds to the most optimistic case for clumping via streaming instability.

\section{Methods}
\label{sec:methods}
To understand whether the action of streaming instability is consistent with recent ALMA observations, we characterise the system observable quantities before and after the action of streaming instability. Our method can be described as a four-step process: (1) we perform hydrodynamics simulations of systems where streaming instability takes place; (2) we define a physical disc model, which allows us to translate the simulation results to  physical systems; (3) we compute the radiative properties of these systems and (4) we compare two observable properties to those derived from ALMA observations by \citet{Tazzari+2020a,Tazzari+2020b}.

\subsection{Numerical simulations}
To simulate the action of streaming instability, we perform 2D hydrodynamics simulations of dust and gas using the hybrid code {\sc Athena}, which treats the gas as a fluid on an Eulerian grid \citep{Stone+2008} and the dust as superparticles on that grid \citep{Bai&Stone2010_athenaparticles}. We use a 2D (vertical and radial directions) shearing box approach, which allows to simulate a portion of the disc located at an arbitrary radius $R_0$ and rotating at angular velocity $\Omega_{\rm k}(R_0)$.

In the following the simulation parameters are given in code units: the time unit is the dynamical time $\Omega_{\rm k}^{-1}$ and the length unit is the gas layer thickness $H_{\rm g}$. The choice of the mass unit is arbitrary, because the equations describing gas and dust interaction (\equationautorefname~\ref{eq:EulerGas},~\ref{eq:particles}) depend on the dust to gas mass ratio, but are independent of the disc mass.

For all the 12 simulation we performed, we considered a box of size $L_r\times L_z =0.05\times 0.15$ ($L_r$ and $L_z$ are the radial and vertical box length, respectively), split in $N_r\times N_z=256\times 768$ cells. In each box we include $N_{\rm tot}=7\cdot 10^5$ particles, equally distributed in $N_{\rm species}=28$ particle species. The Stokes numbers of the different species are uniformly distributed in log space between a minimum and maximum Stokes number, $\tau_{\rm min}$ and $\tau_{\rm max}$; 6 of our simulations are characterised by $\tau_{\rm min}-\tau_{\rm max}=10^{-4}-10^{-1}$, and 6 are characterised by $\tau_{\rm min}-\tau_{\rm max}=10^{-3}-1$. The pressure support parameter is $\Pi=0.025$ in all the simulations, while two values for the dust to gas mass ratio are considered ($Z=0.02$ and $Z=0.03$). 

All the simulations evolve for at least $1500$ dynamical times, in some cases simulations had not reached a steady state in $1500\ \Omega^{-1}$, thus the running time has been extended to $2000\ \Omega^{-1}$ (see \appendixautorefname~\ref{appendix:convergence} for more details).

We use reflecting boundary conditions in the vertical direction and periodic boundary conditions in the radial direction.

As initial conditions, all the particle species are distributed according to a Gaussian shaped distribution in the vertical direction, centred in the disc midplane, whose standard deviation (corresponding to the initial layer thickness) is $H_{\rm d}=0.015\ H_{\rm g}$. We use the parameter $q$ to specify the particle mass distribution per logarithmic particle radius bin
\begin{equation}
    \frac{{\rm d}M_{\rm p}}{{\rm d}\log a} \propto a^{4-q},
    \label{eq:massdistrq}
\end{equation}
or, equivalently, the number distribution $n(a)$
\begin{equation}
    n(a) = \frac{{\rm d}N_{\rm p}}{{\rm d} a} = n_0 a^{-q},
    \label{eq:numberdistrq}
\end{equation}
where $M_{\rm p}$ and $N_{\rm p}$ are the particle mass and number, respectively. In our simulations we consider $q=3$, $q=3.5$, $q=4$ for each choice of $Z$ and set of Stokes numbers.

In \tableautorefname~\ref{tab:parameters} we summurise the parameter used in our simulations.

\begin{table}
    \centering
    \caption{Parameters of the simulations.}
	\label{tab:parameters}
    \begin{tabular}{lccccc}
        \hline
        {Name} & $\tau_{\rm s}^{\rm min}$ & $\tau_{\rm s}^{\rm max}$ & $\Pi$ & $Z$ & $q$\\ 
        \hline
        T41-Z02-Q3 & $10^{-4}$  & $10^{-1}$  & $0.025$  & $0.02$  & $3$\\
        T41-Z02-Q35 & $10^{-4}$  & $10^{-1}$  & $0.025$  & $0.02$  & $3.5$\\
        T41-Z02-Q4 & $10^{-4}$  & $10^{-1}$  & $0.025$  & $0.02$  & $4$\\
        T41-Z03-Q3 & $10^{-4}$  & $10^{-1}$  & $0.025$  & $0.03$  & $3$\\
        T41-Z03-Q35 & $10^{-4}$  & $10^{-1}$  & $0.025$  & $0.03$  & $3.5$\\
        T41-Z03-Q4 & $10^{-4}$  & $10^{-1}$  & $0.025$  & $0.03$  & $4$\\
        T30-Z02-Q3 & $10^{-3}$  & $1$  & $0.025$  & $0.02$  & $3$\\
        T30-Z02-Q35 & $10^{-3}$  & $1$  & $0.025$  & $0.02$  & $3.5$\\
        T30-Z02-Q4 & $10^{-3}$  & $1$  & $0.025$  & $0.02$  & $4$\\
        T30-Z03-Q3 & $10^{-3}$  & $1$  & $0.025$  & $0.03$  & $3$\\
        T30-Z03-Q35 & $10^{-3}$  & $1$  & $0.025$  & $0.03$  & $3.5$\\
        \hline
    \end{tabular}
\end{table}

\subsection{Physical model}
\label{subsec:PhysicalModel}
Studying the optical properties of simulated systems requires the definition of a disc model, allowing conversion of the dimensionless simulation results into physical units. As reference model, we consider the Minimum Mass Solar Nebula (MMSN) model by \citet{Chiang&Youdin2010}.

\citet{Chiang&Youdin2010} model is characterised by a flaring index $f=2/7$, thus the aspect ratio at the generic disc radius $R$ is
\begin{equation}
    H_{\rm g}(R)=H_{\rm g}(R_0) R\left(\frac{R}{R_0}\right)^{2/7},
\end{equation}
where $R_0$ is the box location (in the following we take $R_0=35\rm\ AU$, unless otherwise specified) and the aspect ratio at $R_0$ corresponds to the length unit $L_{\rm unit}=H_{\rm g}(R_0)$.

The time unit is defined as the inverse Keplerian velocity at the box location $t_{\rm unit}=\Omega_{\rm k}^{-1}(R_0)$. Thus, it depends only on the box location and the star mass, which we assume to be the same as that of the Sun $M_*=M_{\rm \odot}$. 

We then choose the arbitrary mass unit by fixing the gas density at $1$ AU and computing the gas density at box location $R_0$ as
\begin{equation}
    \Sigma_{\rm g}(R_0)=\Sigma_{\rm g}({\rm 1AU})\left(\frac{R_0}{\rm AU}\right)^{-p},
    \label{eq:SigmaGas}
\end{equation}
where $p=1.5$ in \citet{Chiang&Youdin2010} model. We do not fix a single value for the gas density, but we vary it in the range $\Sigma_{\rm g}(1\ {\rm AU})= 100-3000\ \rm g/cm^2$ (cf $2200\ \rm g/cm^2$ in \citealt{Chiang&Youdin2010} model), which corresponds to $\Sigma_{\rm g}(R_0)= 0.5-14.5\ \rm g/cm^2$; for each simulation, therefore, we obtain a population of discs characterised by different masses. For a disc characterised by inner and outer radii $R_{\rm in}=0.1\ \rm AU$ and $R_{\rm in}=70\ \rm AU$, respectively, the chosen density distribution corresponds to discs of masses between $10^{-3}\ \rm M_{\odot}$ and $3\cdot10^{-2}\ \rm M_{\odot}$.

Finally, we define the temperature profile
\begin{equation}
    T(R)=T_0\left(\frac{R}{\rm AU}\right)^{-3/7}\left(\frac{L_*}{L_{\odot}}\right)^{1/4},
\end{equation}
where the temperature at 1 AU is $T_0=120$ K (unless stated otherwise) and the temperature profile is the same as that used in the review by \citet{Chiang&Youdin2010}, while the star luminosity is $L_*=L_{\odot}$. The corresponding temperature at the box location ($35\ \rm AU$) for the chosen temperature profile is $T(R_0)\sim 26\ \rm K$.

\subsection{Optical properties}
\label{subsec:OpticalProperties}
To study the dust optical properties, we start by computing the opacity of dust grains; we use \citet{Birnstiel2018} dust code {(which applies the Mie theory combined with optical datasets to compute the system optical properties)}, assuming (similarly to \citealt{Tazzari+2016}) spherical compact grains, whose composition includes water, silicates (from \citealt{Warren&Brandt2008,Draine2003}), troilite (from \citealt{Henning&Stognienko1996}) and organics (from \citealt{Zubko+1996}). Note that the chosen composition is similar to that labelled as `Zubko' in \citet{Birnstiel2018}.\footnote{See \sectionautorefname~\ref{subsec:opacityDiscussion} for a discussion on the composition choice.}

Through \citet{Birnstiel2018} code we obtain the dust material density $\rho_{\rm p}=1.632\ \rm g/cm^3$, used to compute the dust grain size
\begin{equation}
    a = {\frac{2}{\pi}}\frac{\Sigma_{\rm g}\tau_{\rm s}}{\rho_{\rm p}}.
\end{equation}
Then, we extract from \citet{Birnstiel2018} code the opacity corresponding to each simulated grain size and selected wavelength, obtaining the single grain opacity $k_{\nu}^{\rm single}(a)$.

Since the real grain size distribution is expected to be continuous, each simulated $a_i$ represents a set of grain sizes between $a_i$ and $a_i+da_i$, thus we compute the opacity of each grain size as the average opacity
\begin{equation}
    k_{\nu}^{\rm abs}(a_i) = \frac{\bigintss_{a_i-da_i}^{a_i}k_{\nu}^{\rm single}(a)m(a)n(a)da}{\bigintss_{a_i-da_i}^{a_i}m(a)n(a)da}.
    \label{eq:opacity_interp_a}
\end{equation}
Then we compute the size averaged opacity, which gives the system average response (see the \appendixautorefname~\ref{appendix:opacity} for further details)
\begin{equation}
    k_{\nu}^{\rm avg}=\frac{{\bigintss}_{a_{\rm min}}^{a_{\rm max}}k_{\nu}^{\rm abs}(a)m(a)n(a)da}{\bigintss_{a_{\rm min}}^{a_{\rm max}}m(a)n(a)da},
    \label{eq:size_avg_opacity}
\end{equation}
where $m(a)$ is the mass of the dust grain. The main characteristics of $k_{\nu}^{\rm avg}$ as a function of the maximum grain size $a_{\rm max}$ are the presence of a steep increase when $a_{\rm max}\sim \lambda/(2\pi)$, followed by a decline towards larger $a_{\rm max}$ values.

We use the information on $k_{\nu}^{\rm avg}$ to obtain the optical depth
\begin{equation}
	\tau_{\nu}=\frac{k_{\nu}^{\rm avg}\Sigma_{\rm d}}{\cos{i}},
	\label{eq:optDepth}
\end{equation}
where we assume the inclination angle to be $i=0$, and we use this information to obtain the specific intensity (assuming that the system is in local thermodynamical equilibrium)
\begin{equation}
    I_{\nu}=B_{\nu}(T)(1-{\rm e}^{-\tau_{\nu}}),
    \label{eq:SpecInt}
\end{equation}
where $B_{\nu}$ is the Planck function evaluated at the local temperature $T(R)$ (for the location of the considered box we obtain $T(R_0)\sim 26\ \rm K$). Finally, we compute the flux
\begin{equation}
    F_{\nu}\propto\int I_{\nu}dA.
    \label{eq:flux}
\end{equation}

\subsection{Comparison with observations}
\label{subsec:TheoComparisonObservations}
After studying the systems' optical properties, we compare our simulations to the observations obtained by \citet{Tazzari+2020a,Tazzari+2020b} in the Lupus star forming region. Therefore, we focus on the same plane used in \citet{Tazzari+2020a,Tazzari+2020b} to study their data distribution: the optically thick fraction - spectral index plane.

The optically thick fraction is defined as the ratio between the system flux $F_{\nu}$ and the flux that the system would emit if it was a black body
\begin{equation}
    ff = \frac{F_{\nu}}{F_{\nu,\rm black body}} = \frac{\int I_{\nu}dA}{\int B_{\nu}dA},
    \label{eq:ff}
\end{equation}
where $B_{\nu}$ is the Planck function. Note that if a system were optically thick ($\tau_{\nu}\gg 1$), it would have $I_{\nu}\sim B_{\nu}$ (see \equationautorefname~\ref{eq:SpecInt}) and therefore $ff\sim 1$; on the contrary, an optically thin system ($\tau_{\nu}\ll 1$) would be characterised by $I_{\nu}\sim\tau_{\nu} B_{\nu}$, thus $ff\sim {{\tau_\nu}}$.

The spectral index, instead, measures the frequency dependence of the flux
\begin{equation}
    \alpha = \frac{\partial\log F_{\nu}}{\partial\log\nu}.
    \label{eq:alpha}
\end{equation}
It is also useful to introduce the opacity index (see \appendixautorefname~\ref{appendix:opacity} for further information)
\begin{equation}
    \beta=\frac{\partial\log k_{\nu}^{\rm avg}}{\partial\log\nu},
    \label{eq:opacityIndex}
\end{equation}
which describes the variation of the opacity  with frequency (assuming that $k_{\nu}^{\rm avg}\propto\nu^{\beta}$).
$\beta(a_{\rm max})$ describes how $k_{\nu}^{\rm avg}$ varies with $a_{\rm max}$; we thus expect $\beta$ to have a peak for $a_{\rm max}\sim \lambda/(2\pi)$, i.e. where $k_{\nu}^{\rm avg}$ has a steep increase with grain size. Since the value of $a_{\rm max}$ corresponding to peak opacity scales with the wavelength, if we consider any two wavelengths $\lambda_1$ and $\lambda_2$, so that $\lambda_1/(2\pi) < a_{\rm max} < \lambda_2/(2\pi)$, the opacity is much higher at wavelength $\lambda_1$ and hence the value of beta is large (see \appendixautorefname~\ref{appendix:opacity} for further details and plot regarding the spike in $\beta$).

In the limit of low optical depth $F_{\nu}\propto \tau_{\nu}B_{\nu}$, $\alpha$ can be related to the opacity index $\beta$
\begin{equation}
    \alpha = \frac{\partial\log B_{\nu}}{\partial\log \nu} + \beta,
    \label{eq:alpha-beta}
\end{equation}
which can be derived by using \equationautorefname~\ref{eq:alpha} and \equationautorefname~\ref{eq:opacityIndex}. In the Rayleigh-Jeans limit, $\alpha=2+\beta$.  Thus  higher $\alpha$ is associated with higher $\beta$ in the optically thin case; for optically thick emission in the Rayleigh-Jeans limit,   $\alpha= 2$.

We populate the $ff-\alpha$ plane with the simulation results and, by adding on the same plane the data distribution by \citet{Tazzari+2020a,Tazzari+2020b}, we verify whether or not the simulations are consistent with observations. Since the data are available in ALMA bands 3 (100 GHz, 3.3 mm), 6 (230 GHz, 1.33 mm), and 7 (345 GHz, 0.88 mm), we compute the optical properties in these bands; specifically, we compute the optically thick fraction in band 6 $ff_{\rm B6}$, while we use band 3 and band 7 to compute the spectral index $\alpha_{\rm B3-B7}$ (note that we need two values to compute $\alpha$, because it is defined as a derivative).

\section{Clump formation via streaming instability}
\label{sec:density}
\begin{figure*}
    \centering
    \includegraphics[width=1\linewidth]{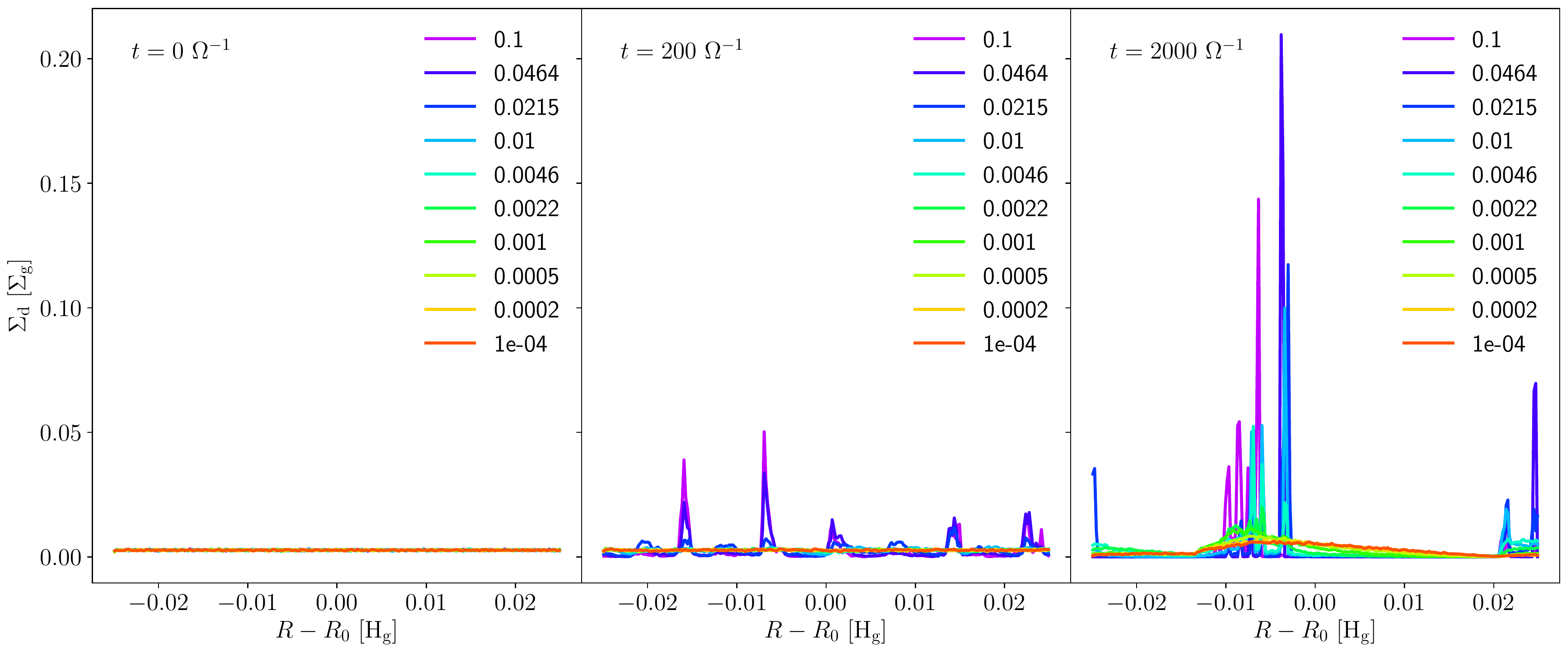}
    \caption{Dust radial density profiles as a function of radius in the simulated box for simulation T41-Z03-Q4. The rainbow colour palette goes from red, which corresponds to the smallest particle species, to magenta, which corresponds to the biggest particle species. The three plots correspond to three different simulation timesteps: the initial condition $t=0\ \rm \Omega^{-1}$ (left panel), {when all the particle species have an uniform density $\Sigma_{\rm d}\sim 0.003$}; an intermediate snapshot $t=200\ \rm \Omega^{-1}$ (central panel); and the result at the end of the simulation $t=2000\ \rm \Omega^{-1}$ (right panel).}
    \label{fig:DensitySpecies}
\end{figure*}
Considering, as an example, simulation T41-Z03-Q4 (see \tableautorefname~\ref{tab:parameters}) in this section we study the properties of clumps formed in the simulated box due to the action of streaming instability.

In \figureautorefname~\ref{fig:DensitySpecies} we show the vertically integrated density profiles of dust ($\sigma_{\rm dust}(R,z)$ is the 2D dust density, in vertical and radial directions)
\begin{equation}
    \Sigma_{\rm dust}(R)=\int_{-L_z/2}^{L_z/2}\sigma_{\rm dust}(R,z)dz,
\end{equation}
for a selection of 10 particle species from the 28 species used in the simulation. The three panels correspond to different timesteps of the system evolution: $t=0\ \Omega^{-1}$ (left panel), $t=200\ \Omega^{-1}$ (central panel) and $t=2000\ \Omega^{-1}$ (right panel). The density profile of each particle species corresponds to a different colour in a rainbow palette, from the smallest particle species in red, to the biggest one in magenta.\footnote{Given the very small area occupied by the clumps at the end of the simulation, they are expected to contribute negligibly to the disc total flux; thus (as anticipated in the previous section) the inclusion of self-gravity is not expected to effect significantly the observable properties.}

The initial density profiles (left panel) are uniform for all the particle species, and clumps are gradually formed as the system evolves (central and right panels). By comparing the central and right panels, we notice that the biggest particle species (magenta, violet and blue lines) are already involved in clumps after 200 $\Omega^{-1}$, while smaller species (cyan and green lines) require more time to participate in clumps; the smallest particle species (yellow and orange lines), instead, never participate in clumping. In fact, particles characterised by Stokes number close to 1 are the most affected by the drag forces, thus they are expected to clump via streaming instability more rapidly; while the smallest particles are highly coupled to the gas, thus they follow gas evolution and do not clump.

\section{Impact of streaming instability on observations: local model}
\label{sec:LocalModelSI}
 In this section we study the observable properties of the simulated 
boxes undergoing streaming instability. In the following, unless specified otherwise, we adopt the disc physical model described in \sectionautorefname~\ref{subsec:PhysicalModel}.

\subsection{Distribution in the \texorpdfstring{$ff-\alpha$}{[ff-alpha]} plane before and after particle clumping}
\label{subsec:distr_beforeafterSI}
\begin{figure*}
    \centering
    \includegraphics[width=1\linewidth]{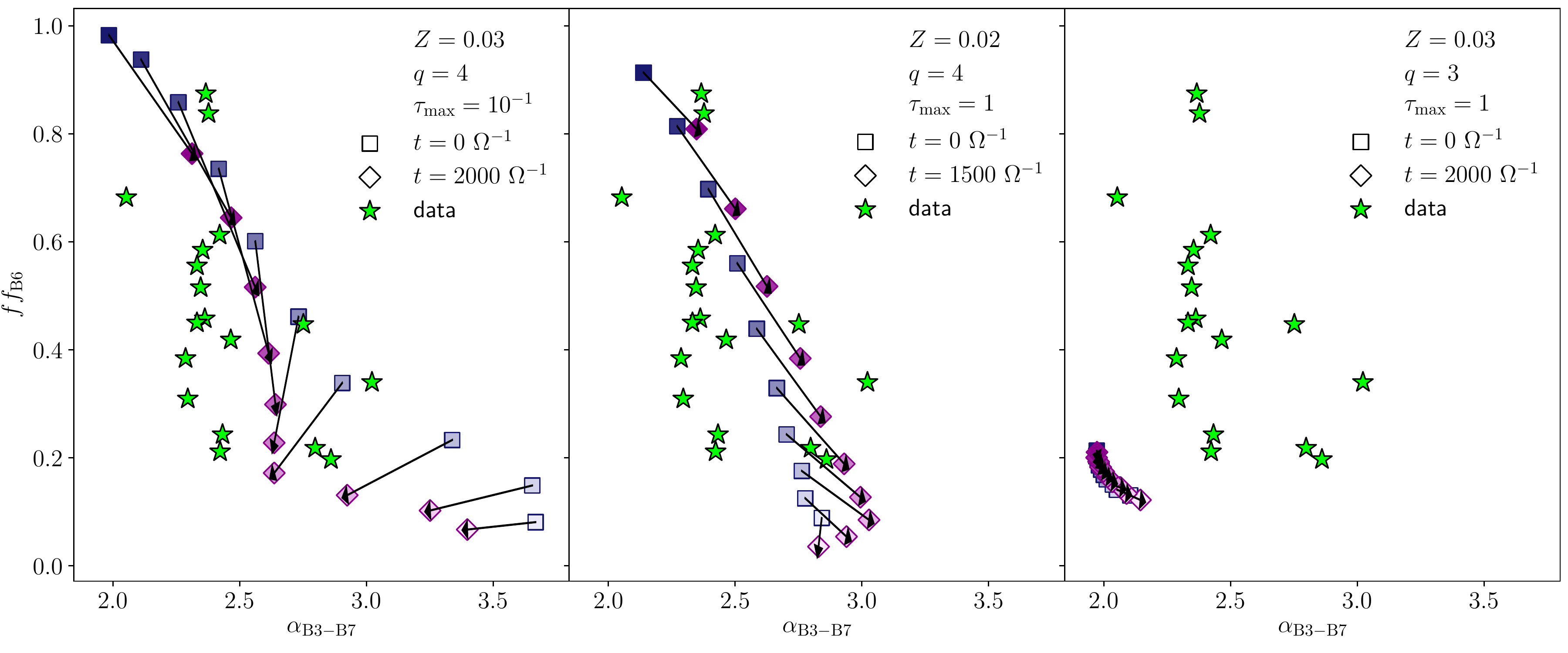}
    \caption{Distribution in the $ff-\alpha$ plane of simulations T41-Z03-Q4 (left panel), T30-Z02-Q4 (central panel), T30-Z03-Q3 (right panel). The blue squares correspond to the distribution for the systems initial conditions, while the magenta diamonds correspond to the distribution at the end of the system evolution; the arrows link the initial $ff-\alpha$ values to the corresponding one at the end of the simulation. The shades of colours corresponds to different gas densities: from the highest density ($\Sigma_{\rm g}(R_0)=14.5\ \rm g/cm^2$, corresponding to $M_{\rm disc}\sim 3\cdot 10^{-2}\ \rm M_{\odot}$), represented by the darkest blue/magenta markers, to the lowest density ($\Sigma_{\rm g}(R_0)=0.5\ \rm g/cm^2$, corresponding to $M_{\rm disc}\sim 10^{-3}\ \rm M_{\odot}$), represented by the lightest blue/magenta markers. The green stars illustrate the data distribution by \citet{Tazzari+2020a}}.
    \label{fig:LocModel_ObservablesInFin}
\end{figure*}

To analyse the effects on the optical properties caused by clump formation via streaming instability, we examine the optically thick fraction (\equationautorefname~\ref{eq:ff}) and the spectral index (\equationautorefname~\ref{eq:alpha}) of the 
integrated emission from the box, both before 
and after the formation of clumps. Since the simulations are scale free (being characterised by the dimensionless numbers set out in \tableautorefname~\ref{tab:parameters}), we can map a single simulation on to a variety of outcomes through considering a range of gas surface densities in the box; we considered 10 values for $\Sigma_{\rm g}$ at the box location, logarithmically distributed between $0.5\ \rm g/cm^2$ and $14.5\ \rm g/cm^2$.

In \figureautorefname~\ref{fig:LocModel_ObservablesInFin} we show the distribution in the $ff-\alpha$ plane for 3 simulations (T41-Z03-Q4, T30-Z02-Q4, T30-Z03-Q3; see \tableautorefname~\ref{tab:parameters}). In all the three panels the blue squares represent the initial distribution of the selected systems, whereas the magenta diamonds correspond to the distribution of the 10 discs at the end of the simulation, i.e. after clumps have formed. The black arrows link together the initial and final condition of the same system. The intensity of the blue/magenta symbols indicates the gas surface density, with the darkest (lightest) shading indicating the highest (lowest) surface density values. For reference, the data distribution is shown in the same plane with the green stars; we notice that clump formation either drives the models towards the area occupied by the data (left panel), or keeps the models in the data area (central panel), or does not affect significantly the distribution (right panel). See \sectionautorefname~\ref{subsec:LocModel_comparisonData} and \sectionautorefname~\ref{sec:IntDiscModel} for a detailed comparison between simulations and data.

Focusing first on the initial conditions (blue squares), we notice that even if they are different in the three considered simulations (due to the difference in the parameter choices), they show some common features. In all  cases, the system characterised by the highest gas surface density (darkest blue square) represents the highest value for $ff$ and the lowest one for $\alpha$: at fixed metallicity this system has the highest dust surface density and hence optical depth and the value of $\alpha$ is therefore close to $2$ as expected for optically thick emission. Conversely, for lower gas  surface density, the lower optical depth results in systems with lower $ff$ and higher  $\alpha$.

We consider first the evolution of $ff$ when clumps form. Clumping involves depositing some of the grains in regions where the optical depth is higher than it was initially; if the clumps are optically thin then this does not affect the over-all flux produced by the box. However, if the clumps are optically thick then it means that some of the emission from grains in the  uniform initial conditions is now hidden and therefore the over-all flux declines. Consequently clumping in general reduces $ff$, though in the case of very optically thin initial conditions (as in the right hand panel where the high maximum Stokes number and relatively low $q$ means that the emission is dominated by large, low opacity  grains) $ff$ may be hardly affected by clumping.

The effect on $\alpha$ is more subtle and depends on the grain size (and hence optical properties) of the grains that are removed from the uniform background and deposited in optically thick clumps where their emission is concealed. Recalling that it is the largest grains that are  deposited in the clumps, the direction of evolution of $\alpha$ depends on whether these largest grains have higher or lower opacity index, $\beta$, than the rest of the grain population.

\begin{figure*}[t]
    \centering
    \includegraphics[width=0.5\linewidth]{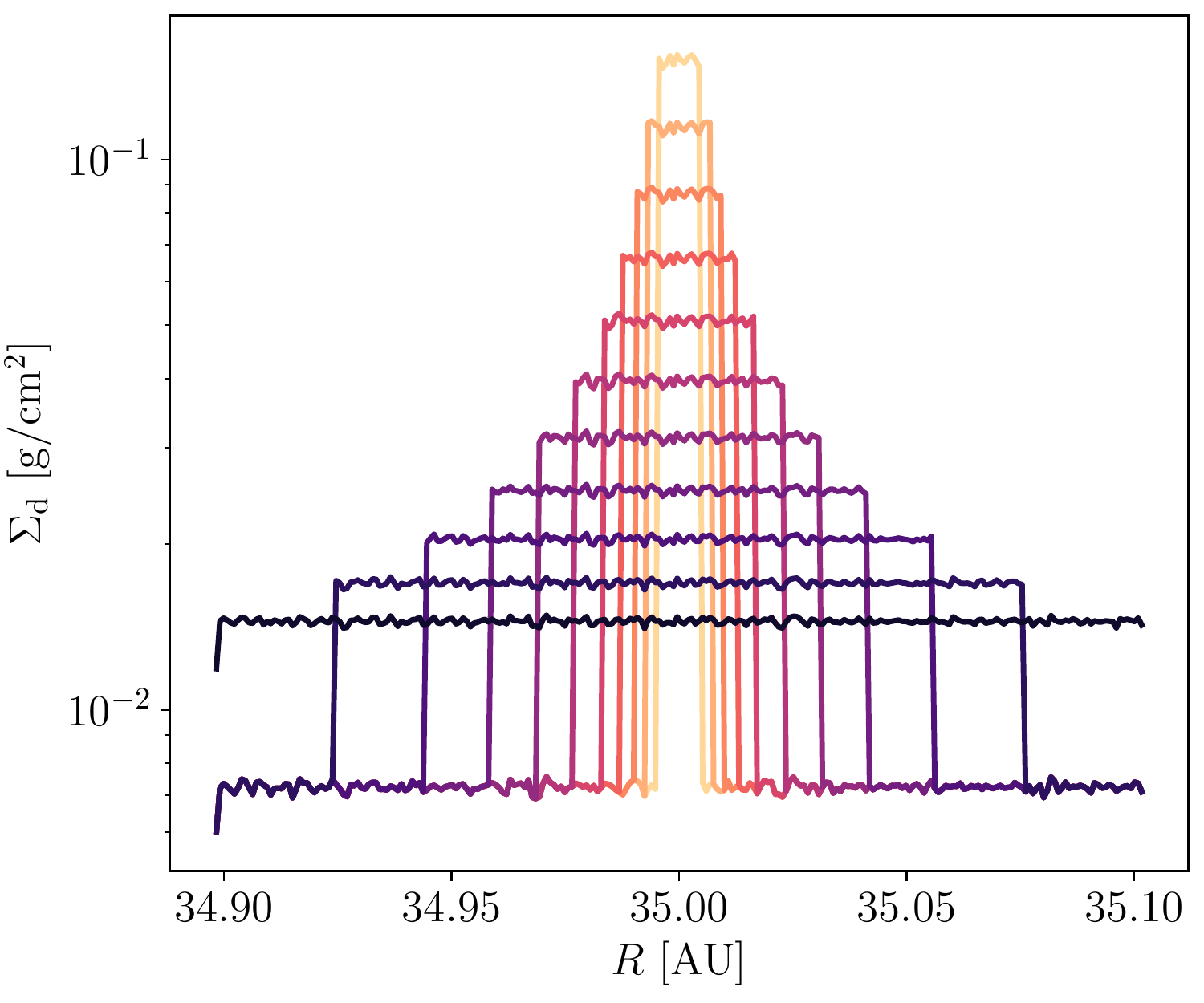}%
    \includegraphics[width=0.5\linewidth]{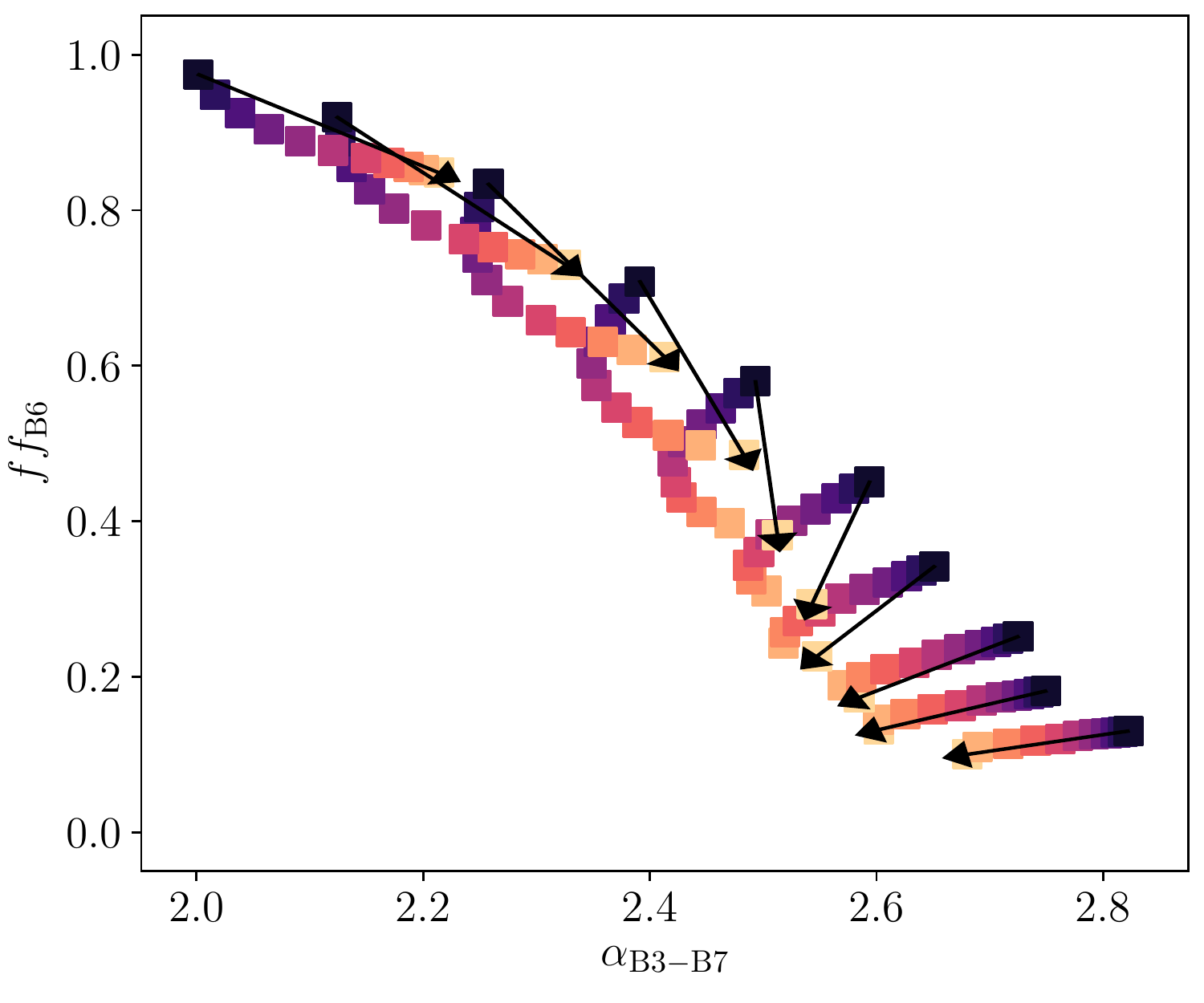}
    \caption{Left panel: re-distribution of the initial homogeneous dust density of simulation T30-Z03-Q4 (for the case with local gas density $\Sigma_{\rm g}=0.5\ \rm g/cm^2$) to create artificial clumps characterised by various heights and widths, assuming that only half of the particle species clump (the biggest ones); a colour code is used, so that dust concentration increases moving from black to pale pink. Right panel: observable quantities for all the density distributions considered in the left panel (the colour of each diamond indicates at which dust density re-distribution that colour corresponds) and different values of $\Sigma_{\rm g}$.}
    \label{fig:ToyModel}
\end{figure*}

We illustrate this effect via a `toy model' (see also \appendixautorefname~\ref{appendix:toymodel} for more details) in which we modify the initial conditions of a system to mimic clump formation by modifying the particles' distribution.
Since in \sectionautorefname~\ref{sec:density} we observed that only particle species characterised by high Stokes numbers participate in clumping, we split the grains in two groups: half of the grains (the smallest ones) are left in the uniform background, whereas the density profile of the remaining grains (the biggest ones) is modified as follows
\begin{equation}
    \Sigma_{j,i}(R)=\begin{dcases}
        \frac{\Sigma_{j,0}}{p_i}\qquad & R_0-dR_i/2 < R < R_0 +dR_i/2,\ j\geq 15\\
        0\qquad & {\rm otherwise}
    \end{dcases}
    \label{eq:ToyModel_sigma}
\end{equation}
where $j$ indicates the $j$-th species; $\Sigma_{j,0}$ is the initial uniform density and $dR_i$ represents the radial fraction of the 2D box in which we concentrate the dust
\begin{equation}
    dR_i = p_i(R_{\rm max}-R_{\rm min}),
    \label{eq:ToyModel_dR}
\end{equation}
where $p_i\in [0,1]$. Note that we are re-distributing only the biggest particles, thus $\Sigma_{\rm d}$ will not be 0 outside the main peak,  since the smallest particles are retained in the background; thus the total dust density for each clump model $i$ is given by
\begin{equation}
    \Sigma_{\rm d,i}(R)=\sum_{j\leq 14}\Sigma_{j,0}(R)+\sum_{j \geq 15}\Sigma_{j,i}(R).
\end{equation}
In the left panel of \figureautorefname~\ref{fig:ToyModel} we show the dust density profiles obtained by redistributing the initial uniform dust density (black line) of simulation T30-Z03-Q4, considering $\Sigma_g=0.5\ \rm g/cm^2$ and 11 values for parameter $p_{i}$, logarithmically distributed between 0.01 and 1 (the clump height increases from the lowest value in black to the highest value in pale pink).
\begin{figure*}
    \centering
    \includegraphics[width=1\linewidth]{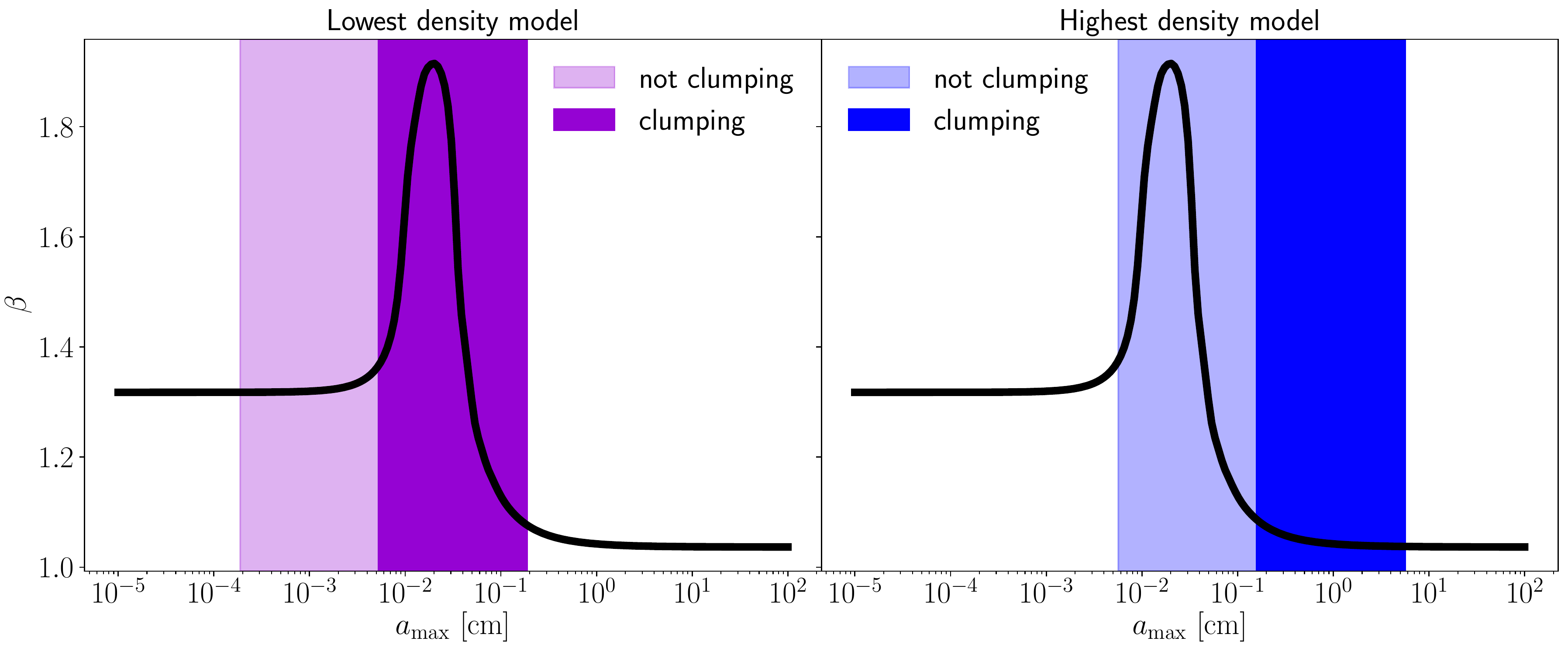}
    \caption{Both panels: opacity index as a function of the grain size obtained using \citet{Birnstiel2018} code (black line); the intense violet/blue area highlights the values of $\beta$ corresponding to the clumping species sizes in the toy model, while the light violet/blue area highlights the the values of $\beta$ corresponding to the not clumping species sizes. The plot in the left panel is obtained considering the lowest density model in this paper ($\Sigma_{\rm g}=0.5\ \rm g/cm^2$), while the plot in the left panel is obtained considering the highest density model in this paper ($\Sigma_{\rm g}=14.5\ \rm g/cm^2$). Note that grain size distribution is assumed to be a power-law with q=4.}
    \label{fig:ToyModel_beta-amax}
\end{figure*}

In the right panel of \figureautorefname~\ref{fig:ToyModel} we show the distribution in the $ff-\alpha$ plane (as before, we consider $\Sigma_{\rm g}=0.5-14.5\ \rm g/cm^2$) obtained by computing the observable quantities for each clump width: the colours in this plot correspond to the dust distribution on the left with the same colour, while the black arrows link together the initial and final conditions of the same system. To explain the behaviour of the spectral index, we focus on the lowest density model (lower right in the plot) and the highest density model (upper left in the plot), which show opposite behaviours in $\alpha$.

We first recall that the spectral index (in the optically thin limit) can be related to the opacity index as follows
\begin{equation}
    \alpha \sim \frac{\partial\log B_{\nu}}{\partial\log \nu} + \beta,
\end{equation}
thus we expect $\alpha$ to increase (decrease) when $\beta$ increases (decreases). We therefore show in \figureautorefname~\ref{fig:ToyModel_beta-amax} the opacity index $\beta$ as a function of the maximum grain size $a_{\rm max}$ (black line). The two panels correspond to the highest and lowest density models on the left and right respectively. In both the panels, two coloured areas highlight the $\beta$ corresponding to clumping (intense violet/blue area) and non-clumping (light blue/violet area) particle species in the simulation (separated at a Stokes number of $\tau_{\rm s}=0.036$, which corresponds to different grain size in the two panels).

We notice that, in the lowest density case, the clumping grains are close to the opacity resonance and hence have higher $\beta$ values than the rest of the population. Once these grains go into the clumps (which are optically thicker than the uniform background), their contribution to the system's emission is downweighted. Therefore the overall $\beta$ decreases and thus we expect $\alpha$ to decrease - as, indeed, happens in the lowest density model in the right panel of \figureautorefname~\ref{fig:ToyModel}. On the contrary, when we consider the opacity index for the highest density model (right panel in \figureautorefname~\ref{fig:ToyModel_beta-amax}), we observe that clumping species are here characterised by lower $\beta$ values than the non-clumping species. Thus we expect both $\beta$ and $\alpha$ to increase after particle clumping; which is consistent with the behaviour of the highest density model in the right panel of \figureautorefname~\ref{fig:ToyModel}.

These behaviours can be related to \figureautorefname~\ref{fig:LocModel_ObservablesInFin}.\footnote{Note that if the clumping factors $p$ for each species are computed from the simulation, then the toy model is able to recover the precise behaviour of simulations (see \appendixautorefname~\ref{appendix:toymodel}).} In the central panel, the high maximum Stokes number implies relatively large grains and hence clumping increases the spectral index  (though this effect is weaker at the lowest gas surface densities where the grain size is smaller at fixed Stokes number). In the left hand panel, conversely, the lower maximum Stokes number places the largest grains in the simulation close to the opacity resonance. Clumping therefore reduces $\alpha$ in this case.

\subsection{Comparison with data}
\label{subsec:LocModel_comparisonData}
\begin{figure*}
    \centering
    \includegraphics[width=1\linewidth]{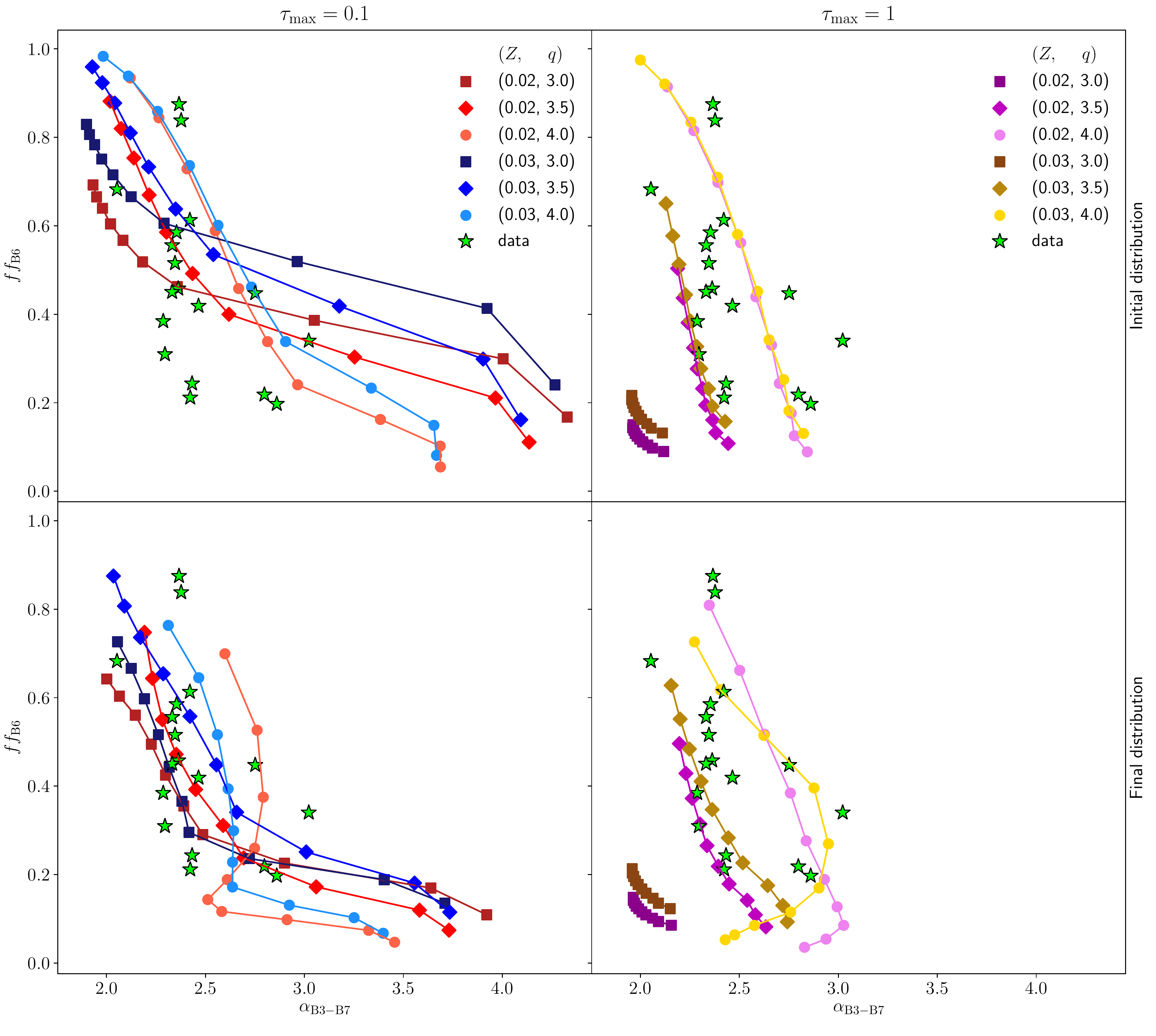}
    \caption{Distribution in the $ff-\alpha$ plane of all the simulation performed (see \tableautorefname~\ref{tab:parameters}). The left panels illustrate the distribution in the observable plane for simulations characterised by $\tau_{\rm max}=0.1$, at the beginning (upper panel) and at the end (lower panel) of the system evolution. The two panels on the right show the same distributions, but for simulations characterised by $\tau_{\rm max}=1$. The values of the main parameters are indicated in the legend, according to the following rule: the squares correspond to $q=3$, the diamonds to $q=3.5$, the dots to $q=4$; in the left panel, the shades of red (blue) are used for systems characterised by $Z=0.02$ ($Z=0.03$); in the right panel, the shades of pink (brown) are used for systems characterised by $Z=0.02$ ($Z=0.03$). In all the panels, the data distribution is illustrated through the green stars.}
    \label{fig:LocModel_ComparisonData}
\end{figure*}
In \figureautorefname~\ref{fig:LocModel_ComparisonData} we show the distributions in the $ff-\alpha$ plane for all the simulated systems and for the data distribution (green stars) obtained by \citet{Tazzari+2020a,Tazzari+2020b}.

We split the simulations into two groups characterised by the same Stokes number range: in the left panels we show the initial (upper panel) and final (lower panel) distribution for simulations characterised by $\tau_{\rm s}=10^{-4}-10^{-1}$; the right panels show the initial (upper panel) and final (lower panel) distribution for simulations characterised by $\tau_{\rm s}=10^{-3}-1$. {In each panel, the different colours refer to different combinations of $q$ and $Z$, as indicated in the plot legend.}

It is worth noticing that the initial distributions (upper panels) partially cover the data distribution. However, most cases characterised by $\tau_{\rm max}=10^{-1}$ have too high optically thick fractions in the highest density models, which are unable to cover the lower left area where data lie; secondly, the lowest density models present too high values for $\alpha$. Similarly, some of the $\tau_{\rm max}=1$ models present $ff$ higher than those of the data (see, in particular, the yellow and pink models).

If we then compare the initial distributions to the final ones (lower panels), we note that the action of streaming instability pushes the simulated systems into the region occupied by data. Indeed, the optically thick fraction is lowered by clump formation, and this effect enables the high density models to re-cover the low $ff$ data that previously were not matched by simulations. Moreover, the spectral index of the low density models is significantly reduced for cases with $\tau_{\rm max}=0.1$; this happens because in these models the clumping species are those characterised by $\beta$ values close to the resonance, thus by removing these grains from the background emission, the $\alpha$ value must decrease. Therefore, for these cases, the action of streaming instability changes the system optical properties so that they become consistent with the data; thus the streaming instability can be considered a candidate to explain the data distribution.

We further underline that there are two simulations (T30-Z02-Q3 and T30-Z03-Q3) whose observable quantities barely evolve when clumps form. This behaviour is related to the fact that, in these cases, both the clumping and the not clumping species are relatively big (due to the particular combination of $\tau_{\rm s}$ and $q$), thus the creation of clumps does not alter significantly the system opacity.

\section{Impact of streaming instability on observations: integrated model}
\label{sec:IntDiscModel}
\begin{figure*}
    \centering
    \includegraphics[width=1\linewidth]{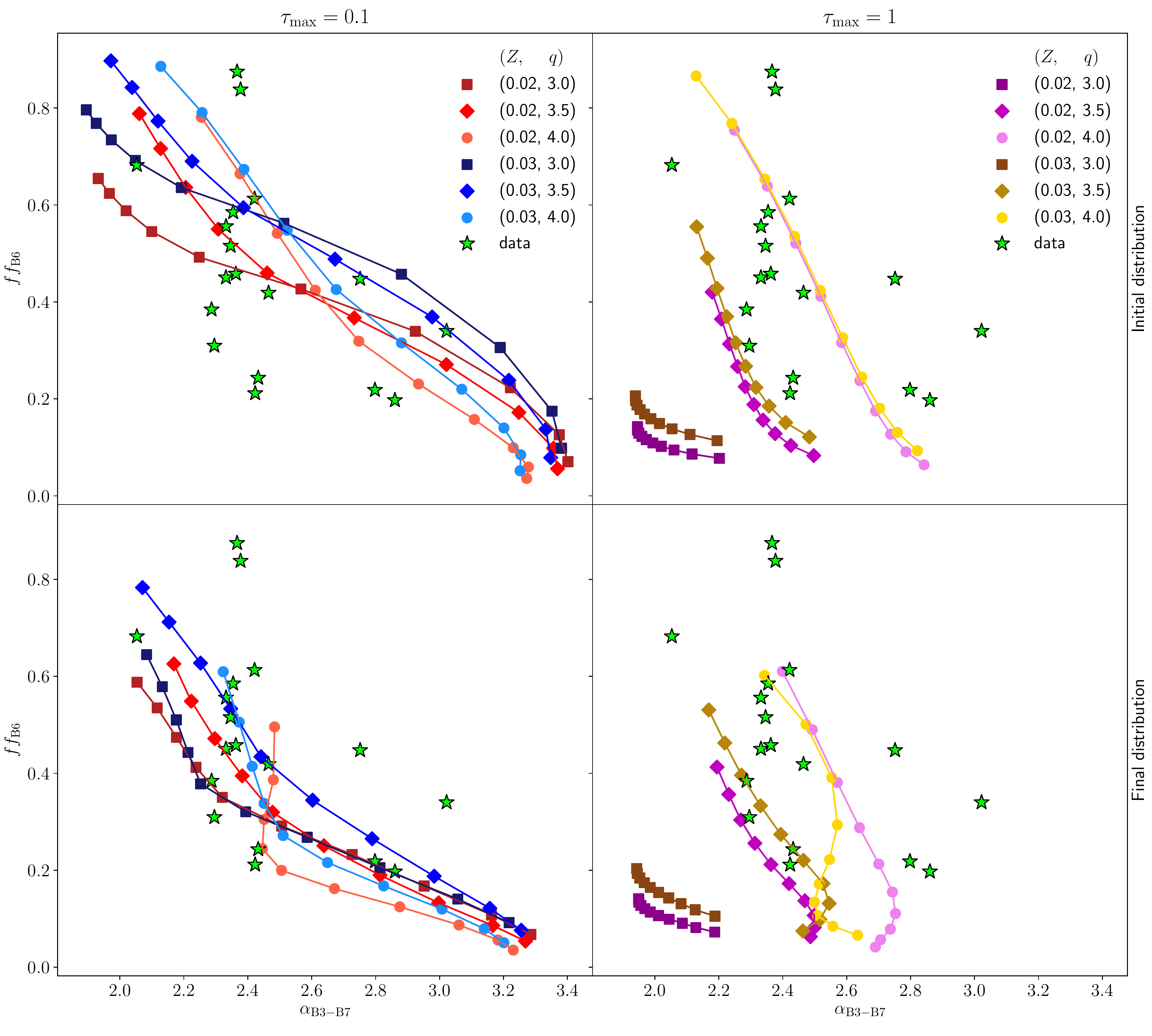}
    \caption{Distribution in the $ff-\alpha$ plane of all the simulation performed (see \tableautorefname~\ref{tab:parameters}) applying the outer disc integrated model. The upper panels show the distribution in the plane for the initial condition, for simulation characterised by $\tau_{\rm max}=0.1$ and $\tau_{\rm max}=1$ in the upper left panel and upper right panel, respectively; the lower panels show the distribution after the system evolution (in the left panel for $\tau_{\rm max}=0.1$ and in the right panel for $\tau_{\rm max}=1$). The values of the main parameters are indicated in the legend (following the same rule as in \figureautorefname~\ref{fig:LocModel_ComparisonData}). In all the panels, the data distribution is illustrated through the green stars.}
    \label{fig:IntModel_ComparisonData}
\end{figure*}
After studying the local observational properties of clumps, we define in this section an integrated disc model to study the global optical properties of systems undergoing streaming instability.

In the integrated disc model we assume azimuthal symmetry, and we define a disc characterised by the following gas density profile
\begin{equation}
    \Sigma_{\rm g}(R)=\Sigma_{\rm g}(1{\rm AU})\left(\frac{R}{\rm AU}\right)^{-p},
    \label{eq:backgroundProfile}
\end{equation}
where $\Sigma_{\rm g}(1{\rm AU})$ varies between $100\ \rm g/cm^2$ and $3000\ \rm g/cm^2$ (note that these values at $1\ \rm AU$ correspond to the values used in the local model at the box location) and we define $R_{\rm in}=0.1\ \rm AU$ and $R_{\rm out}=70\ \rm AU$ as inner and outer radii, respectively. The corresponding dust density profile can be obtained as $\Sigma_{\rm d}=Z\Sigma_{\rm g}$, where $Z$ is chosen according to the considered simulation (see \tableautorefname~\ref{tab:parameters}). {In this model, the Stokes number is assumed to be constant across the disc (found to be a reasonable approximation in the \citealt{Birnstiel+2012} two-population model).}

We divide the defined disc in $N_{\rm rings}=100$ `rings' from $R_{\rm in}=0.1\ \rm AU$ to $R_{\rm out}=70\ \rm AU$, equally spaced on a linear scale.
We then use the $\Sigma_{\rm g}$ and $H_{\rm g}$ local values to map simulations into physical units. If the streaming instability is triggered in a particular ring, then the quantities calculated for that ring are those from the end state of the simulation; otherwise the initial conditions are used.

Once the dust density is obtained, both the opacity and the optical depth are computed at each ring following the same procedure as that used in the local model; in fact, each ring behaves as a single box. In each ring we compute the flux $F_{\nu}^{\rm ring}$, then we sum over all the rings to obtain the total flux
\begin{equation}
    F_{\nu}^{\rm TOT}=\sum_{i=1}^{N_{\rm rings}}F_{\nu}^{\rm ring,i}.
\end{equation}
It is worth noting that for realistic emissivity profiles, the weighting by surface area ensures a relatively large
contribution from the outer disc.

As final step, we compute the observable quantities, which can be simply obtained by using \equationautorefname~\ref{eq:ff} and \equationautorefname~\ref{eq:alpha} where $F_{\nu}^{\rm TOT}$ is used instead of $F_{\nu}$.

To explore different scenarios, we consider 4 different disc models:
\begin{enumerate}
    \item[(1)] all the disc is involved in streaming instability;
    \item[(2)] the inner disc is involved in streaming instability;
    \item[(3)] the outer disc is involved in streaming instability;
    \item[(4)] a ring in the disc is involved in streaming instability.
\end{enumerate}
Following the method outlined above, we determine the distribution in the $ff-\alpha$ plane for all the four integrated disc models. Since the flux is dominated by the outer disc, we found that in models (2) (where the instability occurs only within 35 AU) and (4) (where the instability occurs in a ring of width 10 AU located at 35 AU), the action of streaming instability hardly affects the flux and, therefore, the observable quantities; for the same reason, streaming instability modifies the observable quantities in models (1) and (3), whose behaviour is similar to each other. Therefore, in the following we show the results only for the `outer disc' model, where streaming instability is assumed to take place for $R>R_{\rm SI}$ and we choose $R_{\rm SI}=35\ \rm AU$.

As in the local model, we compare the distribution in the $ff-\alpha$ plane obtained from all the performed simulations to that obtained from the data, and we show the results in \figureautorefname~\ref{fig:IntModel_ComparisonData}. As in \figureautorefname~\ref{fig:LocModel_ComparisonData}, the upper (lower) panels correspond to the initial (final) distributions of system characterised by $\tau_{\rm max}=0.1$ and $\tau_{\rm max}=1$, on the left and on the right, respectively. The legend is the same as that used in \figureautorefname~\ref{fig:LocModel_ComparisonData}.

As in the local model, the initial conditions partially match the data distribution, in fact, the two distributions overlap in the area characterised by $(\alpha,ff)\sim (0.6,2.5)$; nevertheless, the simulated systems are initially unable to cover the central area where  most of the data are located $(\alpha,ff)\sim (0.4,2.5)$. After the action of streaming instability, however, the optically thick fractions decreases, because, as previously explained, the surface density of particles in the optically thin background is reduced by clump formation, reducing the flux and allowing the final distribution to cover the data area that the initial conditions were unable to match. Note that the models do not cover few data located at $(\alpha,ff)\sim(2.4,0.8)$; this can be explained by noticing that $ff$ increases and $\alpha$ decreases as the disc radius is reduced, as shown by Fig. 5 and Fig. 7 in \citet{Tazzari+2020b}. Since our integrated disc models are characterised by a constant disc size ($R_{\rm out}=70\ \rm AU$), we expect the low-$\alpha$ high-$ff$ data to be reproduced by considering smaller discs (indeed, we verified that they are matched by models with $R_{\rm out}=35\ \rm AU$).

It is also worth noticing that in \citet{Tazzari+2020a} the spectral index is correlated with radius, suggesting that the large spectral indices of transition discs can be explained if they have large radii for their masses. This result is consistent with the finding by \citet{Andrews2018} that transition discs are large for their flux; which is also reconcilable with the fact that they have a cavity in their mm-emmision. Such a cavity in the inner disc, however, is not expected to modify significantly the results obtained in this section, as the outer disc is dominant in determining the observable consequences of the action of streaming instability.

We caution that the results might depend on the particular choice of the integrated disc model parameters, see \sectionautorefname~\ref{sec:Parameters_IntModel} for a detailed discussion.

Finally, we underline that if we compute the mm fluxes from our models, we find that they are broadly consistent with those observed, if we adopt the same temperature profile and disc sizes as those observed.

\section{Discussion}
\label{sec:discussion}
\subsection{Influence of model parameters on the distribution (local model)}
\begin{figure}
    \centering
    \includegraphics[width=1\linewidth]{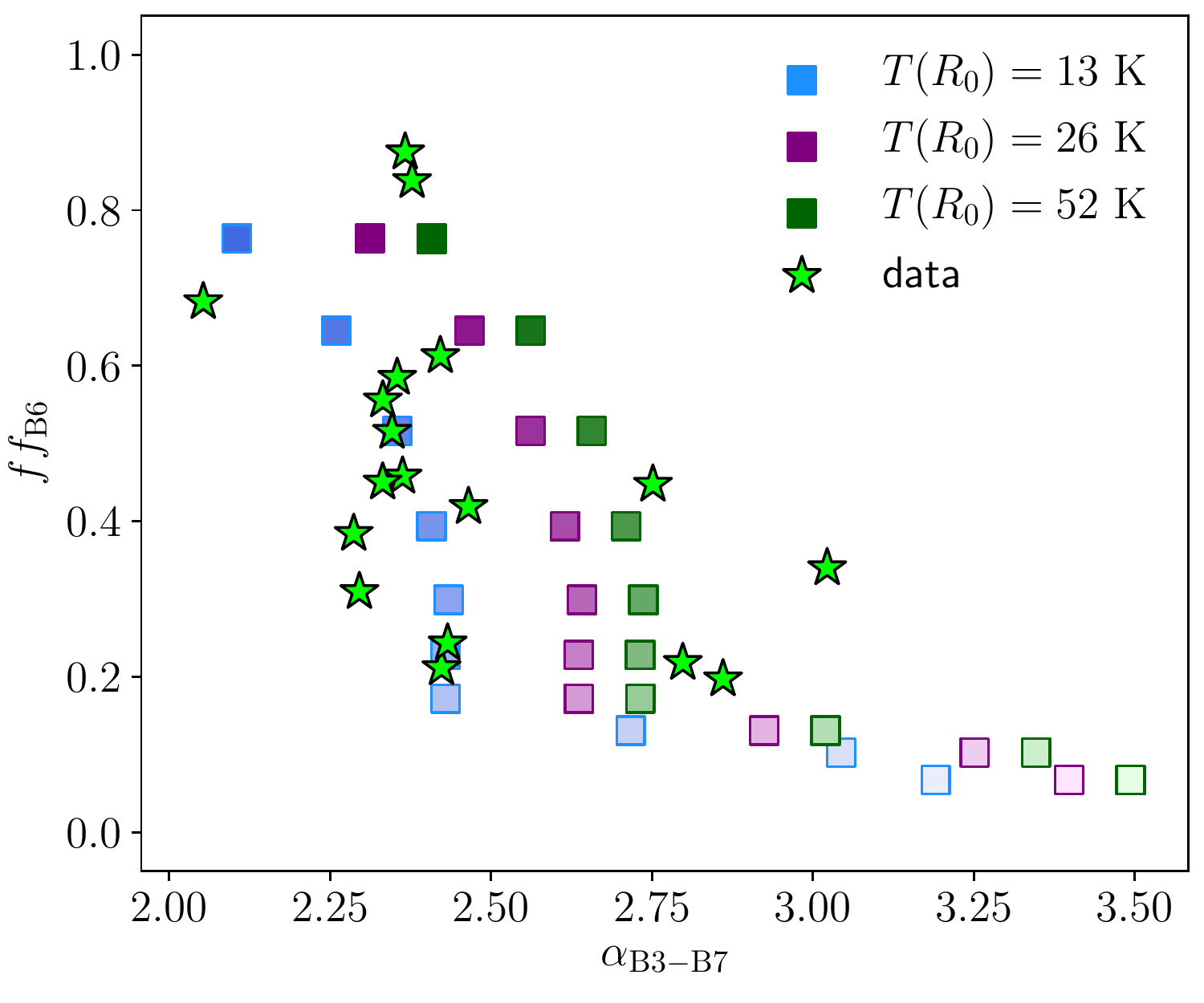}
    \caption{Distribution in the $ff-\alpha$ plane at a box location of 35 AU for simulation T41-Z03-Q4 considering different temperature scaling: $T_0=60\ \rm K$ in blue, $T_0=120\ \rm K$ in purple, $T_0=240\ \rm K$ in green (local temperatures of 13 K, 26 K, 52 K).
    The shades of colours correspond to different values for local $\Sigma_{\rm g}$ (logarithmically distributed in the interval $0.5-14.5\ \rm g/cm^2$). The green stars represent the data distribution by \citet{Tazzari+2020a}}.
    \label{fig:LocModel_InflParameters}
\end{figure}
Although the physics of streaming instability depends only on the parameters chosen in the simulations, the observable quantities also depend on the specific disc model used to convert the simulation dimensionless code units to physical units: the choice of the mass unit influences the grain size, hence the opacity; the local gas density depends on the choice of the box location and the chosen density profile; the temperature profile influences the Planck function.

In \figureautorefname~\ref{fig:LocModel_InflParameters} we test the effect on the distribution in the $ff-\alpha$ plane of changing the temperature, changing $T_0$ to  $60\ \rm K$ in blue, $120\ \rm K$ in purple and  $240\ \rm K$ in green (corresponding to temperatures of 13 K, 26 K, 52 K at the box location.);\footnote{Note that we only calculate the effect of changing temperature on the radiative properties and do not model the effect of changing $\Pi (\propto T^{0.5})$ on the simulations; according to \citet{Bai&Stone2010-2}, a factor two increase of $\Pi$ over the canonical value causes a modest increase in the dust to gas ratio required to trigger the streaming instability.} as an example, we consider simulation T41-Z03-Q4. The shades of each colour represent different local gas density, logarithmically distributed from $\Sigma_{\rm g}=0.5\ \rm g/cm^2$ to $\Sigma_{\rm g}=14.5\ \rm g/cm^2$ (the darker the colour, the higher the density). For comparison, the green stars represent the data distribution.

Changing $T_0$, has no effect on the optically thick fraction, since it only depends on the dust density profile and the opacity (i.e. the grain sizes and their composition), that are independent of $T_0$ (which only modifies the Planck function). Regarding the spectral index (\equationautorefname~\ref{eq:alpha}), we expect it to be unaffected by $T_0$ if the emission is in the Rayleigh-Jeans limit (in which case $\alpha = 2 + \beta$); for low $T_0$, however, the emission is not in the Rayleigh-Jeans regime and this means that $\alpha$ is lower for lower temperatures due to the fact that the higher frequency emission is beyond the Wien peak, resulting in a lower value of $\alpha$.
We see from \figureautorefname~\ref{fig:LocModel_InflParameters} that the effect on the final distribution is relatively small and that the overall distribution is not considerably affected by the choice of $T_0$.

\subsection{Influence of model parameters on the distribution (integrated model)}
\label{sec:Parameters_IntModel}
\begin{figure*}
    \centering
    \includegraphics[width=1\linewidth]{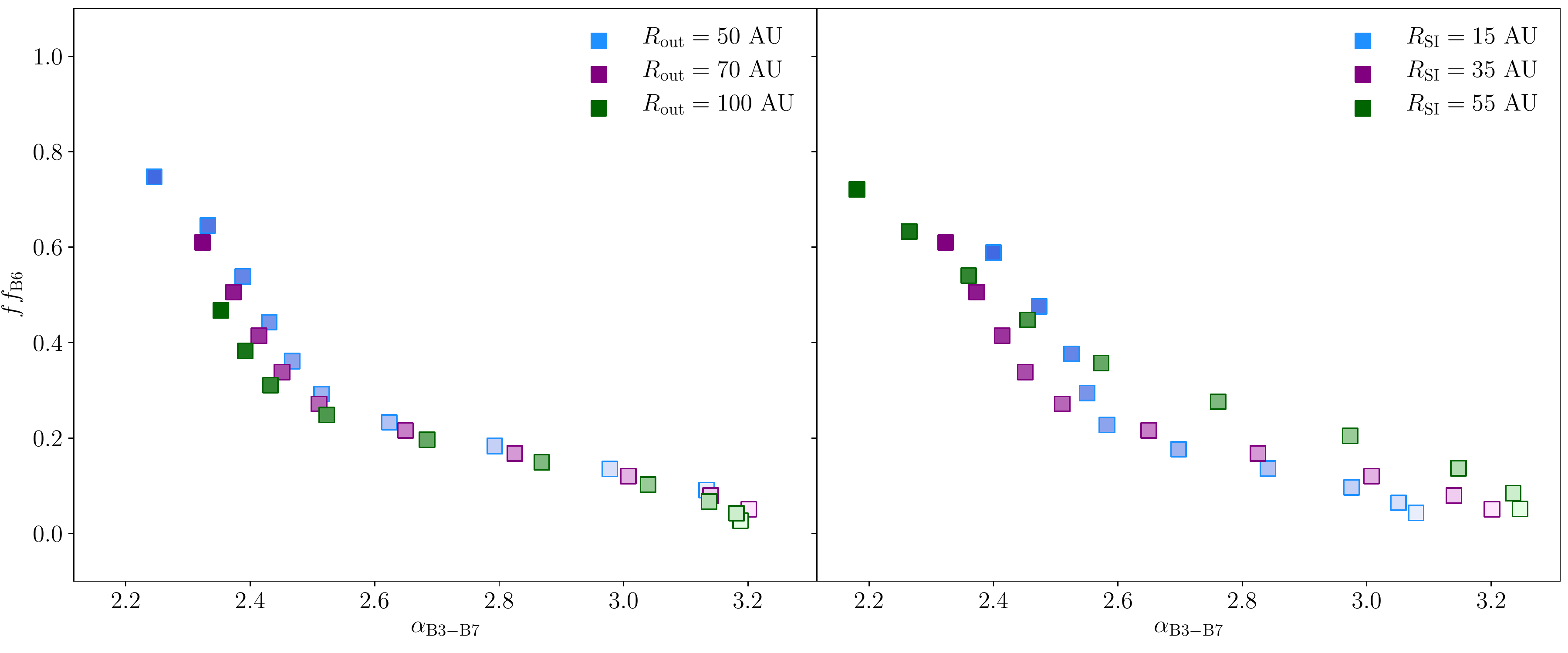}
    \caption{Distributions in the $ff-\alpha$ plane for simulation T41-Z03-Q4, using the `outer disc' integrated model. The left panel shows three distributions obtained using different outer disc radii: $R_{\rm out}=50\ \rm AU$ (blue squares), $R_{\rm out}=70\ \rm AU$ (purple squares), $R_{\rm out}=100\ \rm AU$ (green squares); in all the three cases we rescale $R_{\rm SI}$ so that in all models half of the disc is involved in streaming instability ($R_{\rm SI}=R_{\rm out}/2$). {The range of gas density at $1\ \rm AU$ is the same for all the three cases: $\Sigma_{\rm g}(1{\rm AU})=100-3000\ \rm g/cm^2$.} The right panel shows three distributions obtained by fixing the outer radius $R_{\rm out}=70\ \rm AU$ and changing the radius dividing the area undergoing streaming instability from the area where streaming instability does not take place: blue squares correspond to $R_{\rm SI}=15\ \rm AU$; purple squares to  $R_{\rm SI}=35\ \rm AU$; green squares to  $R_{\rm SI}=55\ \rm AU$. In both the plots, the colours of the squares vary their intensity according to the density of the corresponding disc (the higher is the disc density, the darker is the colour).}
    \label{fig:IntModel_parameters}
\end{figure*}
We also test the influence of the parameter choices in the case of the integrated model, focusing, as an example, on the `outer disc' model. The main parameters to be studied are $R_{\rm out}$ and $R_{\rm SI}$.

In the left panel of \figureautorefname~\ref{fig:IntModel_parameters}, we plot three distributions obtained using the method described in \sectionautorefname~\ref{sec:IntDiscModel} for different disc outer radii: $R_{\rm out}=50\ \rm AU$ (blue squares), $R_{\rm out}=70\ \rm AU$ (purple squares), $R_{\rm out}=100\ \rm AU$ (green squares). The value of $R_{\rm SI}$ is modified so that the portion of disc which is involved in streaming instability is always half of the total disc ($R_{\rm SI}=R_{\rm out}/2$). The considered values for the gas density at $1\ \rm AU$ is the same regardless of the disc outer radius ($\Sigma_{\rm g}(1{\rm AU})=100-3000\ \rm g/cm^2$).
The three obtained distributions are very similar, and the only difference is that smaller discs are slightly optically thicker; in fact, the outer disc portion (which dominates the flux) in smaller discs is optically thicker than in a bigger disc (see \equationautorefname~\ref{eq:backgroundProfile}), therefore overall the disc is expected to be optically thicker.

In the right panel of \figureautorefname~\ref{fig:IntModel_parameters} we test the effect of changing $R_{\rm SI}$ when $R_{\rm out}=70\ \rm AU$ is fixed. In particular, we consider $R_{\rm SI}=15\ \rm AU$ (blue squares), $R_{\rm SI}=35\ \rm AU$ (purple squares), $R_{\rm SI}=55\ \rm AU$ (green squares). 
We notice that when $R_{\rm SI}$, is lower, and thus more of the disc participates in the streaming instability, the optically thick fractions tend to be slightly lower and the bunching towards a smaller range of $\alpha$  (see left hand panel of Figure \ref{fig:LocModel_ObservablesInFin}) also becomes more pronounced.

Overall, the distributions shown in \figureautorefname~\ref{fig:IntModel_parameters} are not significantly influenced by the choice of either $R_{\rm out}$ or $R_{\rm SI}$ in the range considered. Therefore, we can state that the result obtained in \figureautorefname~\ref{fig:IntModel_ComparisonData} is reasonably model independent, provided that the streaming instability is operating over a region of the disc that contributes significantly to the total emission at mm wavelengths.

\subsection{Composition and porosity}
\label{subsec:opacityDiscussion}
In this paper we analysed the optical properties of the simulated systems by considering the composition for dust grains similar to that used in \citet{Tazzari+2016} (which corresponds to the composition labelled as `Zubko' in \citealt{Birnstiel2018}). However, it is important to discuss how the results would be affected by different compositional choices.

We first note, by combining \equationautorefname~\ref{eq:apart} and \equationautorefname~\ref{eq:optDepth}, that at fixed dust to gas ratio, the Stokes number depends on the maximum grain size, optical depth and opacity via:
\begin{equation}
    \tau_{\rm s} \propto \frac{k_{\nu}^{\rm avg}}{\tau_{\nu}} a.
    \label{eq:acomposition}
\end{equation}

We found that, in our models, the streaming instability was  successful in improving the match to  the properties of observed discs on account of the changes in optical depth and spectral index effected when the maximum grain size is close to the opacity resonance. For the compositions adopted here, when the optical depth fraction is several tens of per cent (as in observed systems) and $a$ is close to the opacity resonance, then the Stokes number is in the range where the streaming instability is strong. We can estimate the effect of using, for example, the DSHARP composition (see the case labelled as `default' in \citealt{Birnstiel2018}). From the top panel in Figure 10 of \citet{Birnstiel2018}, we observe that the DSHARP opacity is lower than the `Zubko' one (approximately by a factor of 20) while the $\beta$ peak (middle panel) is located at higher $a$ values (approximately by a factor of 4). This means that for the DSHARP composition  the Stokes number corresponding to the resonance at the same optical depth would be around a factor $5$ less than what we have simulated in this paper. It remains to be seen whether the streaming instability is sufficiently strong at these lower Stokes numbers to have a significant effect on disc radiative properties. 

Likewise we note that if, instead, porous grains are employed, the peak in $\beta$ associated with the opacity resonance virtually disappears (see middle panel of Figure 10 in \citealt{Birnstiel2018}) and thus the streaming instability would have little impact on mm emission from discs. In the absence of the streaming instability, \citet{Tazzari+2020b} showed that  the data in Lupus requires lower $\beta$ values than can be produced by  porous grains. Thus since we find  that the streaming instability does little to the radiative properties of discs in this case, we are led to disfavour the hypothesis of porous grains.

\section{Conclusions}
In this paper we simulated the action of streaming instability by performing 12  2D shearing box simulations with multiple grain sizes  using the {\sc Athena} code. By comparing the results from our simulations to observations in Lupus by \citet{Tazzari+2020a,Tazzari+2020b}, we found that the action of streaming instability is overall consistent with the integrated  emission from discs at mm wavelengths.

Analysis of dust density profiles after clump formation shows that the largest particles participate significantly in clumping, whereas the smallest ones remain  almost uniformly in the background. Streaming
instability thus affects the emission properties of discs in cases where grains towards the upper end of the 
grain size distribution exhibit  a steep dependence of  opacity, and its wavelength dependence, on  grain size. Such a steep dependence  is associated, in the case of compact grains, with  the opacity resonance at a  grain size around a few tenths of a mm. 

We explored the effect of the streaming instability on the location of models in the $ff-\alpha$ (optical depth fraction versus spectral index) plane.
The instability always reduces the optical depth fraction, because the effect of clumping is to reduce the emission from those grains that are translated from optically thin to optically thick regions (in agreement with the finding  by \citealt{Stammler2019} that the streaming 
instability reduces  the system’s optical depth).
The spectral index can evolve in either direction depending on whether the maximum grain size is above or below the size corresponding to the opacity resonance. The net effect is to drive modeled disc systems towards a relatively narrow range of spectral indices (around $\sim 2.5$) which agrees well with the observed distribution of discs in the $ff-\alpha$ plane (see
\figureautorefname~\ref{fig:LocModel_ComparisonData}).

Finally we remark that although we have investigated the  specific scenario of clump generation by  streaming instability, our results are likely to apply, at least qualitatively, to any situation where clumping predominantly affects the largest grains. Clumping is likely strongest for large particles in any effective clumping mechanism; as is the case for trapping by spiral arms in self-gravitating discs (as long as $\tau_{\rm s} \lesssim 1$, \citealt{Booth2016}), pressure maxima created by planets, and vortices. Our primary result (that clumping reduces the optically thick fraction and brings systems to a relatively narrow range of spectral index values) is therefore likely to be of general applicability.

\section*{Acknowledgements}
We thank the referee for his comments that have helped to improve the clarity of the paper, and Marco Tazzari for providing us with the data. CES thanks Peterhouse for a Ph.D. studentship and RAB and CJC acknowledge support from the STFC consolidated grant ST/S000623/1.
This work has also been supported by the European Union's Horizon 2020 research and innovation programme under the Marie Sklodowska-Curie grant agreement No 823823 (DUSTBUSTERS).

\section*{Data availability}
The data underlying this article were provided by \citet{Tazzari+2020a,Tazzari+2020b} by permission. Data will be shared on request to the corresponding author with permission of \citet{Tazzari+2020a,Tazzari+2020b}.



\bibliographystyle{mnras}
\bibliography{references} 

\begin{thebibliography}{}
\makeatletter
\relax
\def\mn@urlcharsother{\let\do\@makeother \do\$\do\&\do\#\do\^\do\_\do\%\do\~}
\def\mn@doi{\begingroup\mn@urlcharsother \@ifnextchar [ {\mn@doi@}
  {\mn@doi@[]}}
\def\mn@doi@[#1]#2{\def\@tempa{#1}\ifx\@tempa\@empty \href
  {http://dx.doi.org/#2} {doi:#2}\else \href {http://dx.doi.org/#2} {#1}\fi
  \endgroup}
\def\mn@eprint#1#2{\mn@eprint@#1:#2::\@nil}
\def\mn@eprint@arXiv#1{\href {http://arxiv.org/abs/#1} {{\tt arXiv:#1}}}
\def\mn@eprint@dblp#1{\href {http://dblp.uni-trier.de/rec/bibtex/#1.xml}
  {dblp:#1}}
\def\mn@eprint@#1:#2:#3:#4\@nil{\def\@tempa {#1}\def\@tempb {#2}\def\@tempc
  {#3}\ifx \@tempc \@empty \let \@tempc \@tempb \let \@tempb \@tempa \fi \ifx
  \@tempb \@empty \def\@tempb {arXiv}\fi \@ifundefined
  {mn@eprint@\@tempb}{\@tempb:\@tempc}{\expandafter \expandafter \csname
  mn@eprint@\@tempb\endcsname \expandafter{\@tempc}}}

\bibitem[\protect\citeauthoryear{{Adachi}, {Hayashi}  \& {Nakazawa}}{{Adachi}
  et~al.}{1976}]{Adachi+1976}
{Adachi} I.,  {Hayashi} C.,   {Nakazawa} K.,  1976, \mn@doi [Progress of
  Theoretical Physics] {10.1143/PTP.56.1756}, \href
  {https://ui.adsabs.harvard.edu/abs/1976PThPh..56.1756A} {56, 1756}

\bibitem[\protect\citeauthoryear{{Andrews} et~al.,}{{Andrews}
  et~al.}{2018}]{Andrews2018}
{Andrews} S.~M.,  et~al., 2018, \mn@doi [\apjl] {10.3847/2041-8213/aaf741},
  \href {https://ui.adsabs.harvard.edu/abs/2018ApJ...869L..41A} {869, L41}

\bibitem[\protect\citeauthoryear{{Ansdell} et~al.,}{{Ansdell}
  et~al.}{2016}]{Ansdell+2016}
{Ansdell} M.,  et~al., 2016, VizieR Online Data Catalog, \href
  {https://ui.adsabs.harvard.edu/abs/2016yCat..18280046A} {p. J/ApJ/828/46}

\bibitem[\protect\citeauthoryear{{Ansdell} et~al.,}{{Ansdell}
  et~al.}{2018}]{Ansdell+2018}
{Ansdell} M.,  et~al., 2018, \mn@doi [\apj] {10.3847/1538-4357/aab890}, \href
  {https://ui.adsabs.harvard.edu/abs/2018ApJ...859...21A} {859, 21}

\bibitem[\protect\citeauthoryear{{Armitage}}{{Armitage}}{2007}]{Armitage2007_BOOK}
{Armitage} P.~J.,  2007, arXiv e-prints, \href
  {https://ui.adsabs.harvard.edu/abs/2007astro.ph..1485A} {pp
  astro--ph/0701485}

\bibitem[\protect\citeauthoryear{{Bai} \& {Stone}}{{Bai} \&
  {Stone}}{2010a}]{Bai&Stone2010_athenaparticles}
{Bai} X.-N.,  {Stone} J.~M.,  2010a, \mn@doi [\apjs]
  {10.1088/0067-0049/190/2/297}, \href
  {https://ui.adsabs.harvard.edu/abs/2010ApJS..190..297B} {190, 297}

\bibitem[\protect\citeauthoryear{{Bai} \& {Stone}}{{Bai} \&
  {Stone}}{2010b}]{Bai&Stone2010-2}
{Bai} X.-N.,  {Stone} J.~M.,  2010b, \mn@doi [\apj]
  {10.1088/0004-637X/722/2/1437}, \href
  {https://ui.adsabs.harvard.edu/abs/2010ApJ...722.1437B} {722, 1437}

\bibitem[\protect\citeauthoryear{{Bai} \& {Stone}}{{Bai} \&
  {Stone}}{2010c}]{Bai&Stone2010}
{Bai} X.-N.,  {Stone} J.~M.,  2010c, \mn@doi [\apjl]
  {10.1088/2041-8205/722/2/L220}, \href
  {https://ui.adsabs.harvard.edu/abs/2010ApJ...722L.220B} {722, L220}

\bibitem[\protect\citeauthoryear{{Birnstiel}, {Klahr}  \&
  {Ercolano}}{{Birnstiel} et~al.}{2012}]{Birnstiel+2012}
{Birnstiel} T.,  {Klahr} H.,   {Ercolano} B.,  2012, \mn@doi [\aap]
  {10.1051/0004-6361/201118136}, \href
  {https://ui.adsabs.harvard.edu/abs/2012A&A...539A.148B} {539, A148}

\bibitem[\protect\citeauthoryear{{Birnstiel}, {Fang}  \&
  {Johansen}}{{Birnstiel} et~al.}{2016}]{Birnstiel+2016}
{Birnstiel} T.,  {Fang} M.,   {Johansen} A.,  2016, \mn@doi [\ssr]
  {10.1007/s11214-016-0256-1}, \href
  {https://ui.adsabs.harvard.edu/abs/2016SSRv..205...41B} {205, 41}

\bibitem[\protect\citeauthoryear{{Birnstiel} et~al.,}{{Birnstiel}
  et~al.}{2018}]{Birnstiel2018}
{Birnstiel} T.,  et~al., 2018, \mn@doi [\apjl] {10.3847/2041-8213/aaf743},
  \href {https://ui.adsabs.harvard.edu/abs/2018ApJ...869L..45B} {869, L45}

\bibitem[\protect\citeauthoryear{{Booth} \& {Clarke}}{{Booth} \&
  {Clarke}}{2016}]{Booth2016}
{Booth} R.~A.,  {Clarke} C.~J.,  2016, \mn@doi [\mnras] {10.1093/mnras/stw488},
  \href {https://ui.adsabs.harvard.edu/abs/2016MNRAS.458.2676B} {458, 2676}

\bibitem[\protect\citeauthoryear{{Brauer}, {Dullemond}  \& {Henning}}{{Brauer}
  et~al.}{2008}]{Brauer+2008}
{Brauer} F.,  {Dullemond} C.~P.,   {Henning} T.,  2008, \mn@doi [\aap]
  {10.1051/0004-6361:20077759}, \href
  {https://ui.adsabs.harvard.edu/abs/2008A&A...480..859B} {480, 859}

\bibitem[\protect\citeauthoryear{{Carrera}, {Johansen}  \& {Davies}}{{Carrera}
  et~al.}{2015}]{Carrera+2015}
{Carrera} D.,  {Johansen} A.,   {Davies} M.~B.,  2015, \mn@doi [\aap]
  {10.1051/0004-6361/201425120}, \href
  {https://ui.adsabs.harvard.edu/abs/2015A&A...579A..43C} {579, A43}

\bibitem[\protect\citeauthoryear{{Chiang} \& {Youdin}}{{Chiang} \&
  {Youdin}}{2010}]{Chiang&Youdin2010}
{Chiang} E.,  {Youdin} A.~N.,  2010, \mn@doi [Annual Review of Earth and
  Planetary Sciences] {10.1146/annurev-earth-040809-152513}, \href
  {https://ui.adsabs.harvard.edu/abs/2010AREPS..38..493C} {38, 493}

\bibitem[\protect\citeauthoryear{{Dominik} \& {Tielens}}{{Dominik} \&
  {Tielens}}{1997}]{Dominik&Tielens1997}
{Dominik} C.,  {Tielens} A.~G.~G.~M.,  1997, \mn@doi [\apj] {10.1086/303996},
  \href {https://ui.adsabs.harvard.edu/abs/1997ApJ...480..647D} {480, 647}

\bibitem[\protect\citeauthoryear{{Dominik}, {Paszun}  \& {Borel}}{{Dominik}
  et~al.}{2016}]{Dominik+2016}
{Dominik} C.,  {Paszun} D.,   {Borel} H.,  2016, arXiv e-prints, \href
  {https://ui.adsabs.harvard.edu/abs/2016arXiv161100167D} {p. arXiv:1611.00167}

\bibitem[\protect\citeauthoryear{{Draine}}{{Draine}}{2003}]{Draine2003}
{Draine} B.~T.,  2003, \mn@doi [\araa]
  {10.1146/annurev.astro.41.011802.094840}, \href
  {https://ui.adsabs.harvard.edu/abs/2003ARA&A..41..241D} {41, 241}

\bibitem[\protect\citeauthoryear{{Dr{\k{a}}{\.z}kowska}, {Alibert}  \&
  {Moore}}{{Dr{\k{a}}{\.z}kowska} et~al.}{2016}]{Drazkowska+2016}
{Dr{\k{a}}{\.z}kowska} J.,  {Alibert} Y.,   {Moore} B.,  2016, \mn@doi [\aap]
  {10.1051/0004-6361/201628983}, \href
  {https://ui.adsabs.harvard.edu/abs/2016A&A...594A.105D} {594, A105}

\bibitem[\protect\citeauthoryear{{Dullemond} et~al.,}{{Dullemond}
  et~al.}{2018}]{Dullemond+2018}
{Dullemond} C.~P.,  et~al., 2018, \mn@doi [\apjl] {10.3847/2041-8213/aaf742},
  \href {https://ui.adsabs.harvard.edu/abs/2018ApJ...869L..46D} {869, L46}

\bibitem[\protect\citeauthoryear{Epstein}{Epstein}{1924}]{Epstein1924}
Epstein P.~S.,  1924, \mn@doi [Phys. Rev.] {10.1103/PhysRev.23.710}, 23, 710

\bibitem[\protect\citeauthoryear{{Garaud}, {Meru}, {Galvagni}  \&
  {Olczak}}{{Garaud} et~al.}{2013}]{Garaud+2013}
{Garaud} P.,  {Meru} F.,  {Galvagni} M.,   {Olczak} C.,  2013, \mn@doi [\apj]
  {10.1088/0004-637X/764/2/146}, \href
  {https://ui.adsabs.harvard.edu/abs/2013ApJ...764..146G} {764, 146}

\bibitem[\protect\citeauthoryear{{Gole}, {Simon}, {Li}, {Youdin}  \&
  {Armitage}}{{Gole} et~al.}{2020}]{Gole+2020}
{Gole} D.~A.,  {Simon} J.~B.,  {Li} R.,  {Youdin} A.~N.,   {Armitage} P.~J.,
  2020, \mn@doi [\apj] {10.3847/1538-4357/abc334}, \href
  {https://ui.adsabs.harvard.edu/abs/2020ApJ...904..132G} {904, 132}

\bibitem[\protect\citeauthoryear{{Henning} \& {Stognienko}}{{Henning} \&
  {Stognienko}}{1996}]{Henning&Stognienko1996}
{Henning} T.,  {Stognienko} R.,  1996, \aap, \href
  {https://ui.adsabs.harvard.edu/abs/1996A&A...311..291H} {311, 291}

\bibitem[\protect\citeauthoryear{{Huang} et~al.,}{{Huang}
  et~al.}{2018}]{Huang+2018}
{Huang} J.,  et~al., 2018, \mn@doi [\apjl] {10.3847/2041-8213/aaf740}, \href
  {https://ui.adsabs.harvard.edu/abs/2018ApJ...869L..42H} {869, L42}

\bibitem[\protect\citeauthoryear{{Johansen} \& {Youdin}}{{Johansen} \&
  {Youdin}}{2007}]{Johansen&Youdin2007}
{Johansen} A.,  {Youdin} A.,  2007, \mn@doi [\apj] {10.1086/516730}, \href
  {https://ui.adsabs.harvard.edu/abs/2007ApJ...662..627J} {662, 627}

\bibitem[\protect\citeauthoryear{{Johansen}, {Oishi}, {Mac Low}, {Klahr},
  {Henning}  \& {Youdin}}{{Johansen} et~al.}{2007}]{Johansen+2007}
{Johansen} A.,  {Oishi} J.~S.,  {Mac Low} M.-M.,  {Klahr} H.,  {Henning} T.,
  {Youdin} A.,  2007, \mn@doi [\nat] {10.1038/nature06086}, \href
  {https://ui.adsabs.harvard.edu/abs/2007Natur.448.1022J} {448, 1022}

\bibitem[\protect\citeauthoryear{{Kobayashi} \& {Tanaka}}{{Kobayashi} \&
  {Tanaka}}{2018}]{Kobayashi&Tanaka2018}
{Kobayashi} H.,  {Tanaka} H.,  2018, \mn@doi [\apj] {10.3847/1538-4357/aacdf5},
  \href {https://ui.adsabs.harvard.edu/abs/2018ApJ...862..127K} {862, 127}

\bibitem[\protect\citeauthoryear{{Kobayashi}, {Tanaka}  \&
  {Okuzumi}}{{Kobayashi} et~al.}{2016}]{Kobayashi+2016}
{Kobayashi} H.,  {Tanaka} H.,   {Okuzumi} S.,  2016, \mn@doi [\apj]
  {10.3847/0004-637X/817/2/105}, \href
  {https://ui.adsabs.harvard.edu/abs/2016ApJ...817..105K} {817, 105}

\bibitem[\protect\citeauthoryear{{Kokubo} \& {Ida}}{{Kokubo} \&
  {Ida}}{1996}]{Kokubo&Ida1996}
{Kokubo} E.,  {Ida} S.,  1996, \mn@doi [\icarus] {10.1006/icar.1996.0148},
  \href {https://ui.adsabs.harvard.edu/abs/1996Icar..123..180K} {123, 180}

\bibitem[\protect\citeauthoryear{{Krapp}, {Ben{\'\i}tez-Llambay}, {Gressel}  \&
  {Pessah}}{{Krapp} et~al.}{2019}]{Krapp+2019}
{Krapp} L.,  {Ben{\'\i}tez-Llambay} P.,  {Gressel} O.,   {Pessah} M.~E.,  2019,
  \mn@doi [\apjl] {10.3847/2041-8213/ab2596}, \href
  {https://ui.adsabs.harvard.edu/abs/2019ApJ...878L..30K} {878, L30}

\bibitem[\protect\citeauthoryear{{Laibe} \& {Price}}{{Laibe} \&
  {Price}}{2014}]{Laibe&Price2014}
{Laibe} G.,  {Price} D.~J.,  2014, \mn@doi [\mnras] {10.1093/mnras/stu1367},
  \href {https://ui.adsabs.harvard.edu/abs/2014MNRAS.444.1940L} {444, 1940}

\bibitem[\protect\citeauthoryear{{Li}, {Youdin}  \& {Simon}}{{Li}
  et~al.}{2018}]{Li+2018}
{Li} R.,  {Youdin} A.~N.,   {Simon} J.~B.,  2018, \mn@doi [\apj]
  {10.3847/1538-4357/aaca99}, \href
  {https://ui.adsabs.harvard.edu/abs/2018ApJ...862...14L} {862, 14}

\bibitem[\protect\citeauthoryear{{Lissauer}}{{Lissauer}}{1993}]{Lissauer1993}
{Lissauer} J.~J.,  1993, \mn@doi [\araa] {10.1146/annurev.aa.31.090193.001021},
  \href {https://ui.adsabs.harvard.edu/abs/1993ARA&A..31..129L} {31, 129}

\bibitem[\protect\citeauthoryear{{Liu} \& {Ji}}{{Liu} \&
  {Ji}}{2020}]{Liu&Ji2020}
{Liu} B.,  {Ji} J.,  2020, \mn@doi [Research in Astronomy and Astrophysics]
  {10.1088/1674-4527/20/10/164}, \href
  {https://ui.adsabs.harvard.edu/abs/2020RAA....20..164L} {20, 164}

\bibitem[\protect\citeauthoryear{{Ormel}, {Dullemond}  \& {Spaans}}{{Ormel}
  et~al.}{2010}]{Ormel+2010}
{Ormel} C.~W.,  {Dullemond} C.~P.,   {Spaans} M.,  2010, \mn@doi [\icarus]
  {10.1016/j.icarus.2010.06.011}, \href
  {https://ui.adsabs.harvard.edu/abs/2010Icar..210..507O} {210, 507}

\bibitem[\protect\citeauthoryear{{Papaloizou} \& {Terquem}}{{Papaloizou} \&
  {Terquem}}{2006}]{Papaloizou&Terquem2006}
{Papaloizou} J. C.~B.,  {Terquem} C.,  2006, \mn@doi [Reports on Progress in
  Physics] {10.1088/0034-4885/69/1/R03}, \href
  {https://ui.adsabs.harvard.edu/abs/2006RPPh...69..119P} {69, 119}

\bibitem[\protect\citeauthoryear{{Pinte} \& {Laibe}}{{Pinte} \&
  {Laibe}}{2014}]{Pinte&Laibe2014}
{Pinte} C.,  {Laibe} G.,  2014, \mn@doi [\aap] {10.1051/0004-6361/201220545},
  \href {https://ui.adsabs.harvard.edu/abs/2014A&A...565A.129P} {565, A129}

\bibitem[\protect\citeauthoryear{{Rafikov}}{{Rafikov}}{2003}]{Rafikov2003}
{Rafikov} R.~R.,  2003, \mn@doi [\aj] {10.1086/345971}, \href
  {https://ui.adsabs.harvard.edu/abs/2003AJ....125..942R} {125, 942}

\bibitem[\protect\citeauthoryear{{Safronov} \& {Zvjagina}}{{Safronov} \&
  {Zvjagina}}{1969}]{Safronov&Zvjagina1969}
{Safronov} V.~S.,  {Zvjagina} E.~V.,  1969, \mn@doi [\icarus]
  {10.1016/0019-1035(69)90013-X}, \href
  {https://ui.adsabs.harvard.edu/abs/1969Icar...10..109S} {10, 109}

\bibitem[\protect\citeauthoryear{{Schaffer}, {Yang}  \& {Johansen}}{{Schaffer}
  et~al.}{2018}]{Schaffer+2018}
{Schaffer} N.,  {Yang} C.-C.,   {Johansen} A.,  2018, \mn@doi [\aap]
  {10.1051/0004-6361/201832783}, \href
  {https://ui.adsabs.harvard.edu/abs/2018A&A...618A..75S} {618, A75}

\bibitem[\protect\citeauthoryear{{Squire} \& {Hopkins}}{{Squire} \&
  {Hopkins}}{2020}]{Squire&Hopkins2020}
{Squire} J.,  {Hopkins} P.~F.,  2020, \mn@doi [\mnras]
  {10.1093/mnras/staa2311}, \href
  {https://ui.adsabs.harvard.edu/abs/2020MNRAS.498.1239S} {498, 1239}

\bibitem[\protect\citeauthoryear{{Stammler}, {Dr{\k{a}}{\.z}kowska},
  {Birnstiel}, {Klahr}, {Dullemond}  \& {Andrews}}{{Stammler}
  et~al.}{2019}]{Stammler2019}
{Stammler} S.~M.,  {Dr{\k{a}}{\.z}kowska} J.,  {Birnstiel} T.,  {Klahr} H.,
  {Dullemond} C.~P.,   {Andrews} S.~M.,  2019, \mn@doi [\apjl]
  {10.3847/2041-8213/ab4423}, \href
  {https://ui.adsabs.harvard.edu/abs/2019ApJ...884L...5S} {884, L5}

\bibitem[\protect\citeauthoryear{{Stone}, {Gardiner}, {Teuben}, {Hawley}  \&
  {Simon}}{{Stone} et~al.}{2008}]{Stone+2008}
{Stone} J.~M.,  {Gardiner} T.~A.,  {Teuben} P.,  {Hawley} J.~F.,   {Simon}
  J.~B.,  2008, \mn@doi [\apjs] {10.1086/588755}, \href
  {https://ui.adsabs.harvard.edu/abs/2008ApJS..178..137S} {178, 137}

\bibitem[\protect\citeauthoryear{{Takeuchi} \& {Lin}}{{Takeuchi} \&
  {Lin}}{2002}]{Takeuchi&Lin2002}
{Takeuchi} T.,  {Lin} D.~N.~C.,  2002, \mn@doi [\apj] {10.1086/344437}, \href
  {https://ui.adsabs.harvard.edu/abs/2002ApJ...581.1344T} {581, 1344}

\bibitem[\protect\citeauthoryear{{Tazzari} et~al.,}{{Tazzari}
  et~al.}{2016}]{Tazzari+2016}
{Tazzari} M.,  et~al., 2016, \mn@doi [\aap] {10.1051/0004-6361/201527423},
  \href {https://ui.adsabs.harvard.edu/abs/2016A&A...588A..53T} {588, A53}

\bibitem[\protect\citeauthoryear{{Tazzari} et~al.,}{{Tazzari}
  et~al.}{2020a}]{Tazzari+2020a}
{Tazzari} M.,  et~al., 2020a, arXiv e-prints, \href
  {https://ui.adsabs.harvard.edu/abs/2020arXiv201002248T} {p. arXiv:2010.02248}

\bibitem[\protect\citeauthoryear{{Tazzari}, {Clarke}, {Testi}, {Williams},
  {Facchini}, {Manara}, {Natta}  \& {Rosotti}}{{Tazzari}
  et~al.}{2020b}]{Tazzari+2020b}
{Tazzari} M.,  {Clarke} C.~J.,  {Testi} L.,  {Williams} J.~P.,  {Facchini} S.,
  {Manara} C.~F.,  {Natta} A.,   {Rosotti} G.,  2020b, arXiv e-prints, \href
  {https://ui.adsabs.harvard.edu/abs/2020arXiv201002249T} {p. arXiv:2010.02249}

\bibitem[\protect\citeauthoryear{{Umurhan}, {Estrada}  \& {Cuzzi}}{{Umurhan}
  et~al.}{2020}]{Umurhan+2020}
{Umurhan} O.~M.,  {Estrada} P.~R.,   {Cuzzi} J.~N.,  2020, \mn@doi [\apj]
  {10.3847/1538-4357/ab899d}, \href
  {https://ui.adsabs.harvard.edu/abs/2020ApJ...895....4U} {895, 4}

\bibitem[\protect\citeauthoryear{{Warren} \& {Brandt}}{{Warren} \&
  {Brandt}}{2008}]{Warren&Brandt2008}
{Warren} S.~G.,  {Brandt} R.~E.,  2008, \mn@doi [Journal of Geophysical
  Research (Atmospheres)] {10.1029/2007JD009744}, \href
  {https://ui.adsabs.harvard.edu/abs/2008JGRD..11314220W} {113, D14220}

\bibitem[\protect\citeauthoryear{{Weidenschilling}}{{Weidenschilling}}{1977}]{Weidenschilling1977}
{Weidenschilling} S.~J.,  1977, \mn@doi [\mnras] {10.1093/mnras/180.1.57},
  \href {https://ui.adsabs.harvard.edu/abs/1977MNRAS.180...57W} {180, 57}

\bibitem[\protect\citeauthoryear{{Yang} \& {Johansen}}{{Yang} \&
  {Johansen}}{2014}]{Yang&Johansen2014}
{Yang} C.-C.,  {Johansen} A.,  2014, \mn@doi [\apj]
  {10.1088/0004-637X/792/2/86}, \href
  {https://ui.adsabs.harvard.edu/abs/2014ApJ...792...86Y} {792, 86}

\bibitem[\protect\citeauthoryear{{Yang}, {Johansen}  \& {Carrera}}{{Yang}
  et~al.}{2017}]{Yang+2017}
{Yang} C.~C.,  {Johansen} A.,   {Carrera} D.,  2017, \mn@doi [\aap]
  {10.1051/0004-6361/201630106}, \href
  {https://ui.adsabs.harvard.edu/abs/2017A&A...606A..80Y} {606, A80}

\bibitem[\protect\citeauthoryear{{Youdin} \& {Goodman}}{{Youdin} \&
  {Goodman}}{2005}]{Youdin&Goodman2005}
{Youdin} A.~N.,  {Goodman} J.,  2005, \mn@doi [\apj] {10.1086/426895}, \href
  {https://ui.adsabs.harvard.edu/abs/2005ApJ...620..459Y} {620, 459}

\bibitem[\protect\citeauthoryear{{Youdin} \& {Johansen}}{{Youdin} \&
  {Johansen}}{2007}]{Youdin&Johansen2007}
{Youdin} A.,  {Johansen} A.,  2007, \mn@doi [\apj] {10.1086/516729}, \href
  {https://ui.adsabs.harvard.edu/abs/2007ApJ...662..613Y} {662, 613}

\bibitem[\protect\citeauthoryear{{Zhu} \& {Yang}}{{Zhu} \&
  {Yang}}{2020}]{Zhu&Yang2020}
{Zhu} Z.,  {Yang} C.-C.,  2020, arXiv e-prints, \href
  {https://ui.adsabs.harvard.edu/abs/2020arXiv200801119Z} {p. arXiv:2008.01119}

\bibitem[\protect\citeauthoryear{{Zubko}, {Mennella}, {Colangeli}  \&
  {Bussoletti}}{{Zubko} et~al.}{1996}]{Zubko+1996}
{Zubko} V.~G.,  {Mennella} V.,  {Colangeli} L.,   {Bussoletti} E.,  1996, in
  {K{\"a}ufl} H.~U.,  {Siebenmorgen} R.,  eds, The Role of Dust in the
  Formation of Stars. p.~333

\makeatother
\end{thebibliography}



\appendix

\section{Convergence}
\label{appendix:convergence}
\begin{figure}
\centering
\includegraphics[width=1\linewidth]{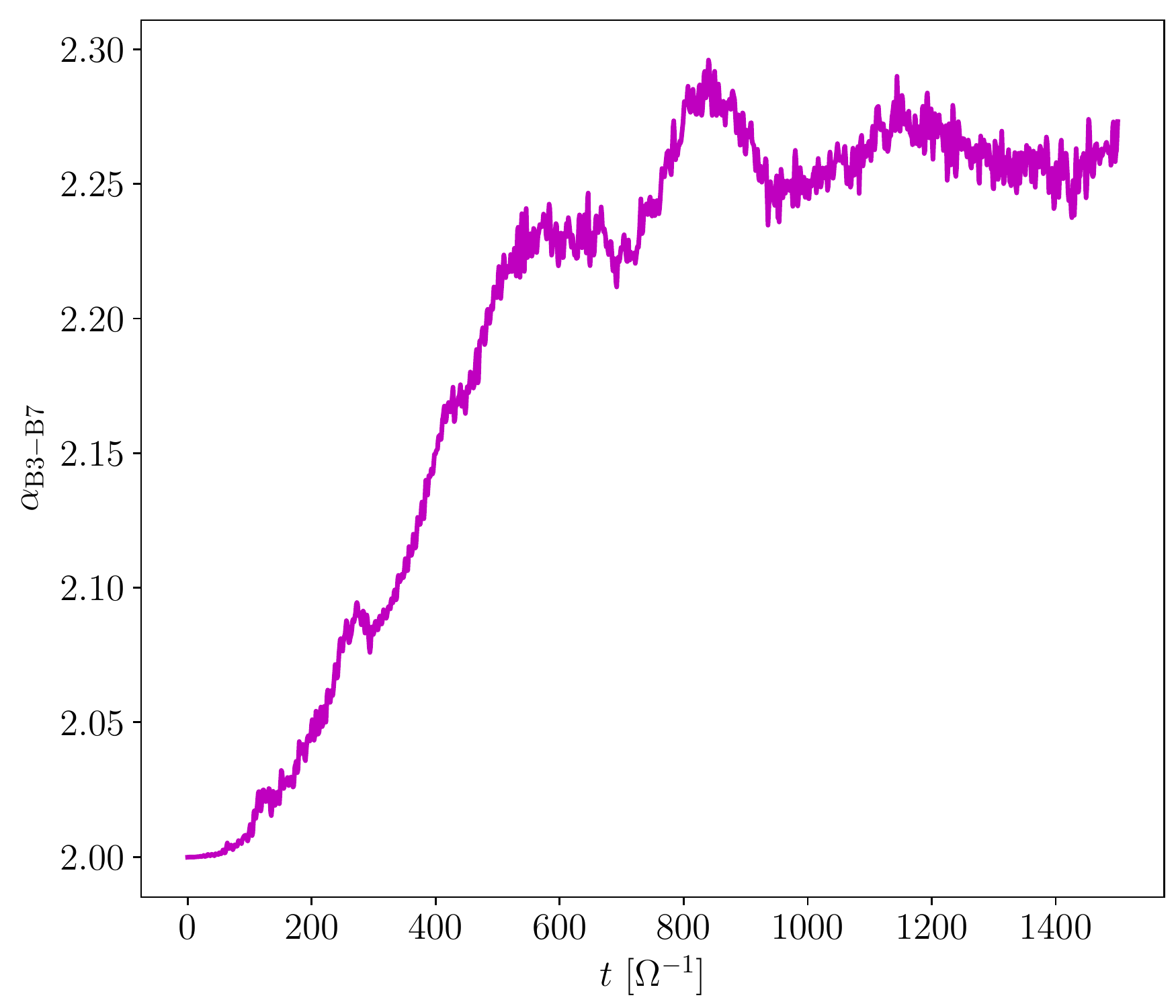}
\caption{Spectral index as a function of time for simulation T30-Z03-Q4, considering the standard disc model with local gas density $\Sigma_{\rm g}=14.5\ \rm g/cm^2$.}
\label{fig:convergenceAlpha}
\end{figure}

Since we are interested in the changes of optical properties before and after the action of streaming instability, we must ensure that at the end of our the running time our simulations have converged. We therefore computed for each simulation the observable quantities as a function of time, and we considered the simulation as converged when $\alpha$ and $ff$ stopped to increase/decrease. In \figureautorefname~\ref{fig:convergenceAlpha} we show, as an example, the variation of the spectral index with time for simulation T30-Z03-Q4: in the first $800$ dynamical times $\alpha$ increases considerably, then it stabilises on a nearly constant value. We performed this convergence test for all the simulations, finding that the most of simulations have converged after $1500\ \rm \Omega^{-1}$, apart from a few simulations (simulations T41-Z02-Q4, T41-Z03-Q4, T30-Z02-Q3, T30-Z03-Q3) which required a longer evolution to reach convergence ($2000\ \rm \Omega^{-1}$).

\section{Opacity}
\label{appendix:opacity}
\begin{figure*}
\centering
\includegraphics[width=0.33\linewidth]{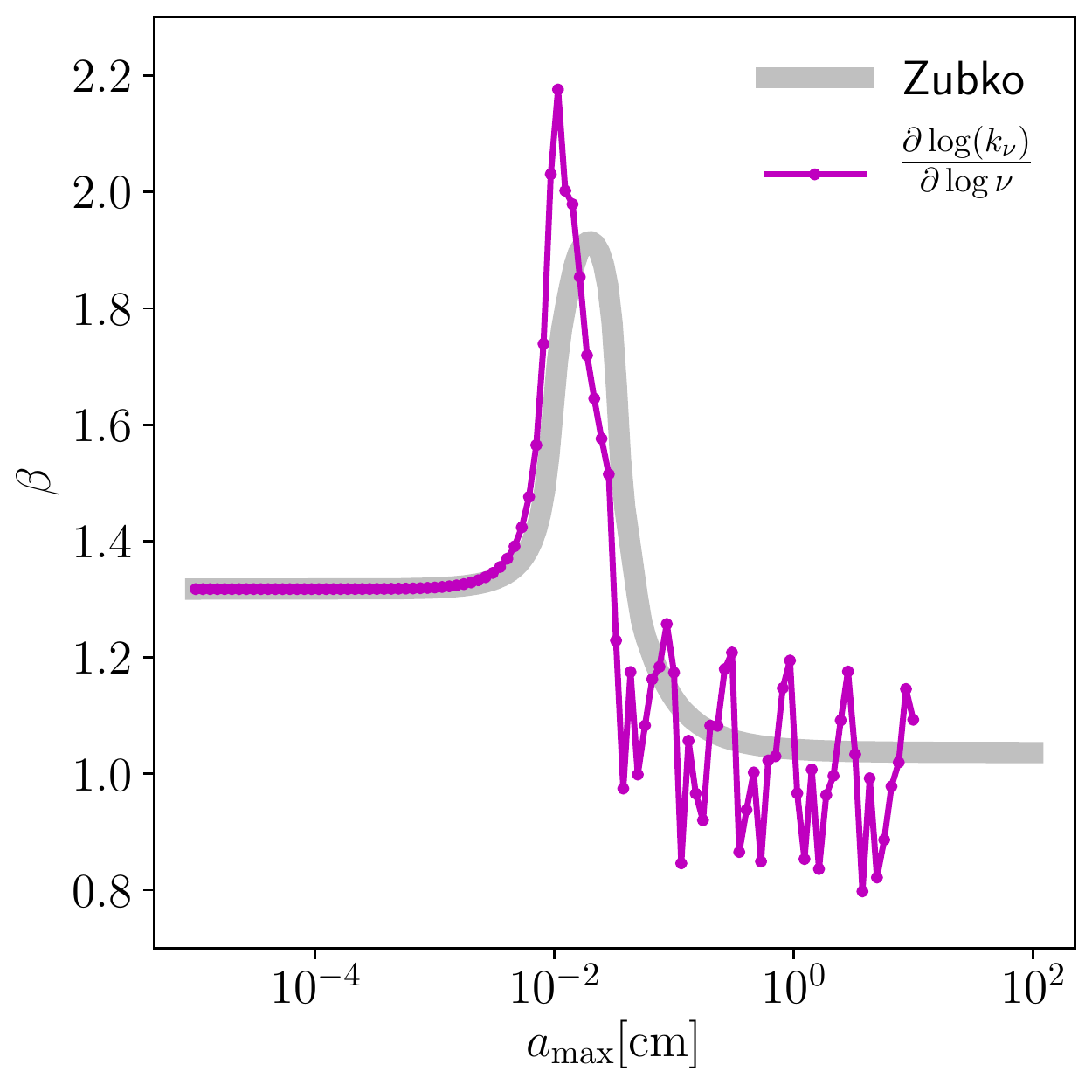}%
\includegraphics[width=0.33\linewidth]{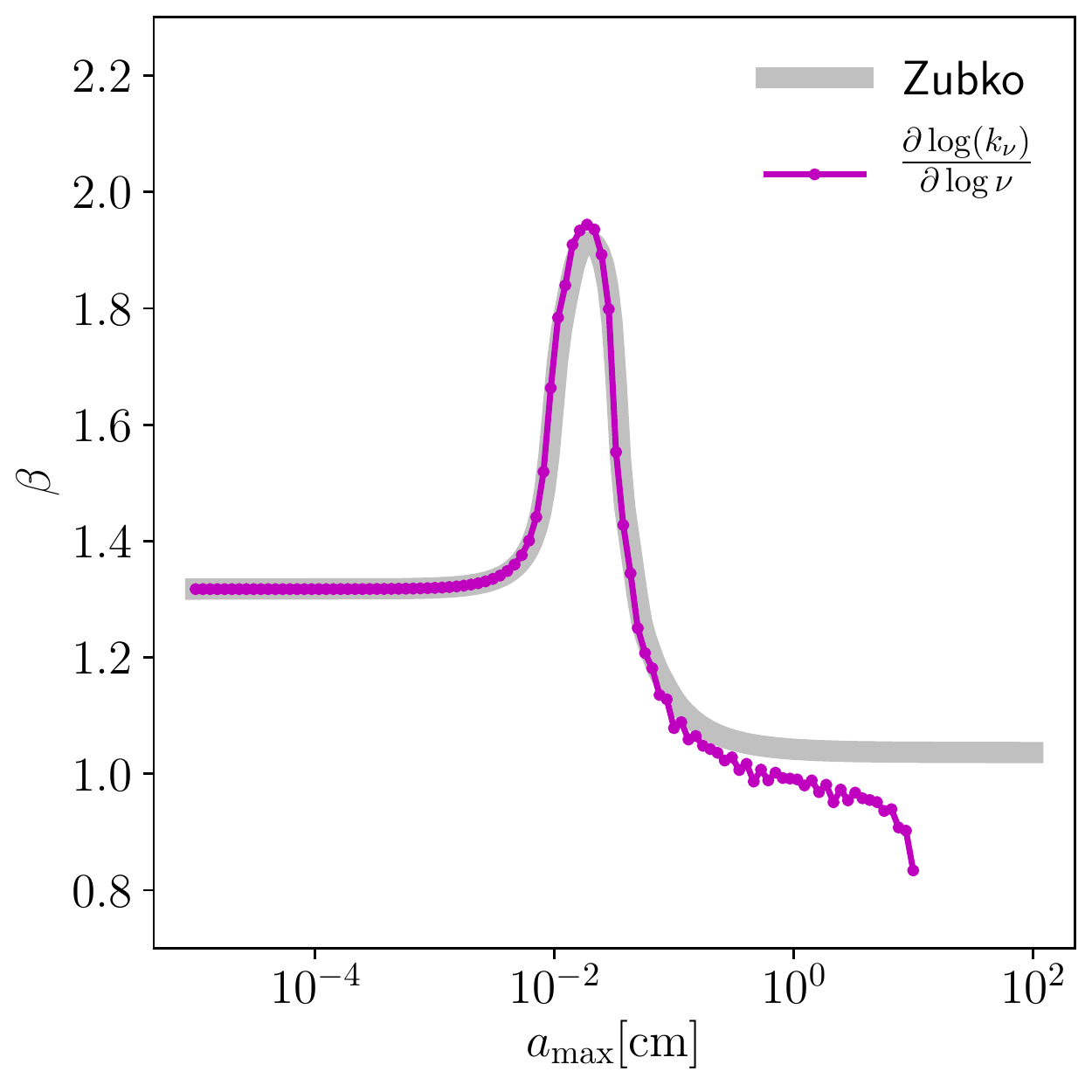}%
\includegraphics[width=0.33\linewidth]{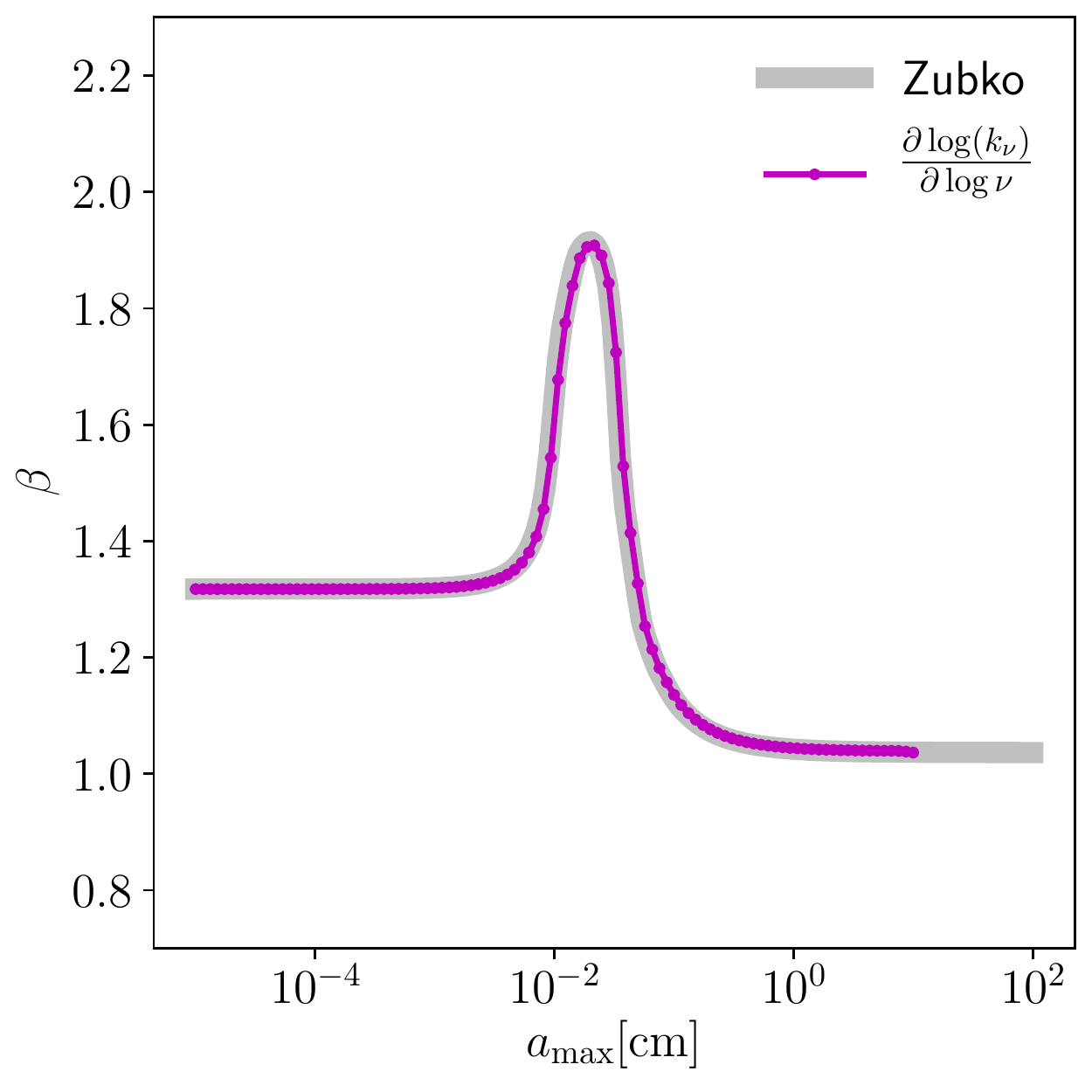}
\caption{Opacity index as a function of the maximum grain size for different models, obtained from simulation T41-Z03-Q4. The grey line is the result obtained from \citet{Birnstiel2018} code; the magenta line is the result of our computation. The left panel shows the result when we consider the single grain opacity and 7 particle species; the middle panel shows the result when we interpolate the grain distribution using 28 particle species; the right panel shows the result when we consider that each grain represents a set of grains and we extrapolate the distribution to include the low mass grains.}
\label{fig:beta}
\end{figure*}
Considering as an example simulation T41-Z03-Q4, we explain here the details of the method applied to compute the size averaged opacity for our simulations. The method can be split in three phases: (1) we compute the single grain opacity, using \citet{Birnstiel2018} code; (2) we interpolate over the grain size interval that each simulated grain represents (\equationautorefname~\ref{eq:opacity_interp_a}); (3) we compute the size averaged opacity (\equationautorefname~\ref{eq:size_avg_opacity}). To test whether our computation works properly we compute the opacity index $\beta$ as a function of the maximum grain size $a_{\rm max}$ and we compare it to the same quantity obtained with \citet{Birnstiel2018} code.

In the left panel of \figureautorefname~\ref{fig:beta}, the magenta line shows $\beta$ obtained by considering a simulation characterised by 7 particle species and without operating the grain interpolation. The thick grey line show the result obtained from \citet{Birnstiel2018} code (as well as in the other two panels). This computation clearly unable to return the correct opacity, as the low number of species and the absence of grain size interpolation causes strong oscillations. By increasing the number of particle species to 28 (central panel), we note a considerable improvement in the computation of opacity, since the majority of oscillations disappear; nevertheless, at high $a_{\rm max}$, $\beta$ seems to be underestimated. Once we introduce also the size grain interpolation (right panel), we are finally able to recover precisely the correct shape of $\beta-a_{\rm max}$. This analysis allowed us to choose the correct number of particle species to insert in our simulation (28 species) and to verify that \equationautorefname~\ref{eq:opacity_interp_a} is needed in order to compute the opacity properly.

\begin{figure}
\centering
\includegraphics[width=0.8\linewidth]{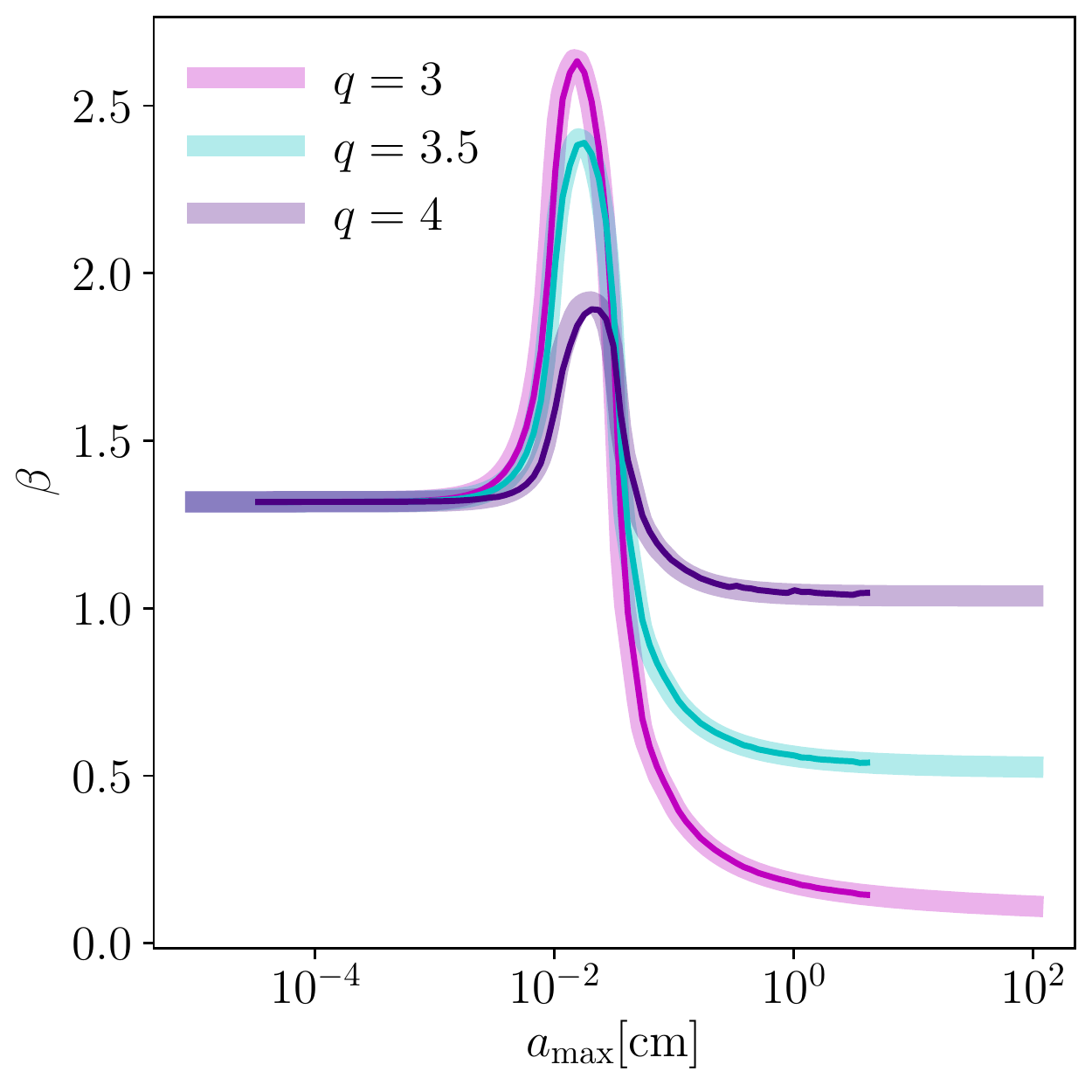}
\caption{Opacity index as a function of the maximum grain size, for different grain size distributions: q=3 (magenta), q=3.5 (cyan), q=4 (violet). The thick translucent lines corresponds to the values obtained through \citet{Birnstiel2018} code; the thin solid lines corresponds to the values computed using our sets of dust grain distributions.}
\label{fig:betaq}
\end{figure}
In \figureautorefname~\ref{fig:betaq} we applied the method outlined above to the cases $q=3$ (magenta line), $q=3.5$ (cyan line) and $q=4$ (violet line). The thin lines refer to our computation, the thick translucent ones refer to the corresponding result obtained through \citet{Birnstiel2018} code. Even though for small $a_{\rm max}$, the three cases behave similarly, they differ significantly among each other for the peak height and the value for big $a_{\rm max}$. In particular, for big grains, the opacity variation with frequency is steeper and steeper as $q$ increases.

\section{Toy model}
\label{appendix:toymodel}
In \sectionautorefname~\ref{subsec:distr_beforeafterSI} we used a toy model to illustrate how the optical properties of clumping particles affect the increase/decrease in $\alpha$. Here, we show that the toy model is effective in reproducing the distribution in the $ff-\alpha$ plane of simulated systems.

\begin{figure*}
    \centering
    \includegraphics[width=1\linewidth]{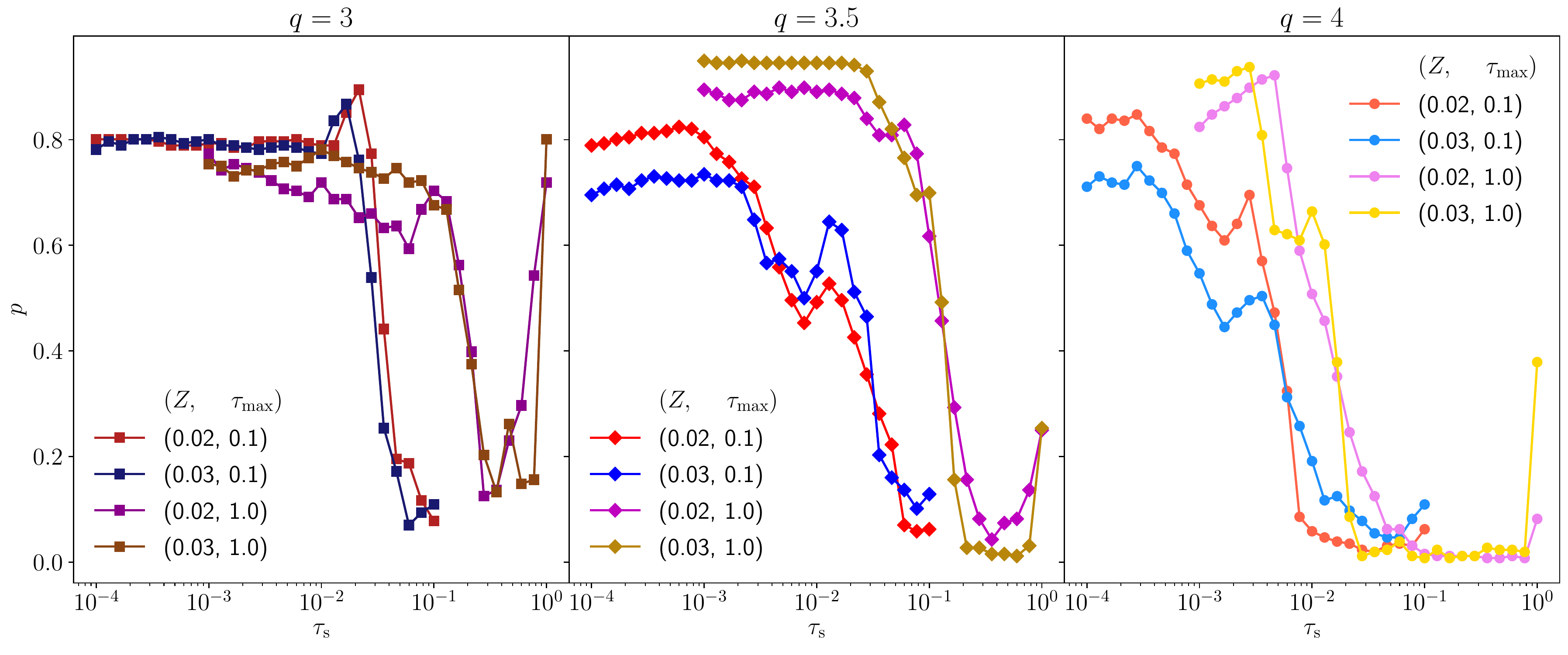}
    \caption{Parameter $p$ as a function of the particle species Stokes number $\tau_{\rm s}$ for all the simulation performed. The results are split into three groups depending on their $q$ parameter: $q=3$ in the left panel, $q=3.5$ in the central panel, $q=4$ in the right panel.}
    \label{fig:toymodel_p_tauS}
\end{figure*}

We compute for each simulation the level of clumping for each particle species the mean density ratio (i.e. we compute the dust density for each particle species and average that over all the particles), then we use the result to identify the clump area as the area where the final mean density ratio is higher than the initial one, and finally we obtain $p$ by using \equationautorefname~\ref{eq:ToyModel_dR}. The result is shown in \figureautorefname~\ref{fig:toymodel_p_tauS}.

\begin{figure*}
    \centering
    \includegraphics[width=1\linewidth]{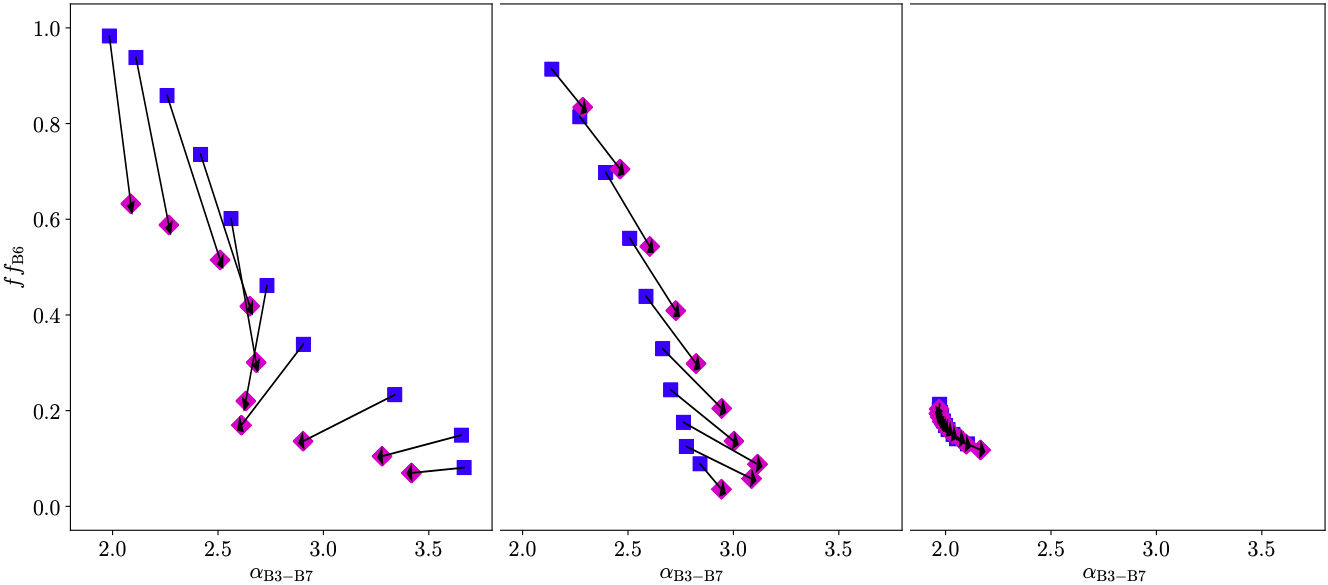}
    \caption{Distribution in the $ff-\alpha$ plane obtained applying the toy model to systems characterised by the same parameters as simulations considered in \figureautorefname~\ref{fig:LocModel_ObservablesInFin}: T41-Z03-Q4 (left panel), T30-Z02-Q4 (central panel), T30-Z03-Q3 (right panel).}
    \label{fig:test_toymodel}
\end{figure*}

We then use the values of $p$ computed above to mimic clump formation of each particle species\footnote{Note that the simplified approach used in \sectionautorefname~\ref{subsec:distr_beforeafterSI}, in which we arbitrarily split the particles in clumping/not-clumping species, is useful to explain the different behaviours of $\alpha$, but it cannot reproduce the distribution in the $ff-\alpha$ plane precisely.} through \equationautorefname~\ref{eq:ToyModel_dR} and \equationautorefname~\ref{eq:ToyModel_sigma} and we compute the corresponding distribution in the $ff-\alpha$ plane. We test the toy model by applying it to the initial conditions of simulations plotted in \figureautorefname~\ref{fig:LocModel_ObservablesInFin} and we show the results in \figureautorefname~\ref{fig:test_toymodel} (the panels, from left to right, correspond to the panels in \figureautorefname~\ref{fig:LocModel_ObservablesInFin}). By comparing \figureautorefname~\ref{fig:test_toymodel} to \figureautorefname~\ref{fig:LocModel_ObservablesInFin}, we can notice that the behaviour of the observable quantities in the toy model is similar to that obtained through the actual simulations; this confirms that both the definition used to obtain $p$ in \figureautorefname~\ref{fig:toymodel_p_tauS} and the method of mimic clump formation by redistributing the particles are effective in reproducing the observable quantities.



\bsp	
\label{lastpage}
\end{document}